\documentclass[11pt]{article}
\usepackage{a4wide}

\usepackage[dvipdfmx,hiresbb]{graphicx} 
\usepackage{mediabb}                     

\usepackage{color}
\definecolor{refkey}{rgb}{1,0,1}
\definecolor{labelkey}{rgb}{1,0,1}

\usepackage{hyperref}

\newcommand{\addedSecond}[1]{{\color{black}#1}}
\newcommand{\Magenta}[1]{{\color{black}#1}}

\newcommand{\al}[1]{\begin{align}#1\end{align}}
\newcommand{\als}[1]{\begin{align*}#1\end{align*}}
\newcommand{\ab}[1]{\left|#1\right|}
\newcommand{\paren}[1]{\left(#1\right)}
\newcommand{\fn}[1]{\!\left(#1\right)}
\newcommand{\sqbr}[1]{\left[#1\right]}
\newcommand{\br}[1]{\left\{#1\right\}}

\newcommand{\MSbar}{\ensuremath{\overline{\text{MS}}} }
\newcommand{\wt}{\widetilde}

\usepackage{amsmath,amssymb}
\newcommand{\df}{\text{d}}

\newcommand{\mc}{\mathcal}
\newcommand{\V}{\mathcal V}

\usepackage{epsf}

\newcommand{\cred}[1]{{\color{black}#1}}
\newcommand{\rev}[1]{{\color{black}#1}}	
\newcommand{\cmag}[1]{{\color{black}#1}}

\newcommand{\GeV}{\ensuremath{\,\text{GeV} }}

\newcommand{\nn}{\nonumber\\}

\begin{document}

\title{\vbox{
\baselineskip 14pt
\hfill \hbox{\normalsize KUNS-2461, OU-HET/792}
} \vskip 1cm
\bf \Large Minimal Higgs inflation \vskip 0.5cm
}
\author{
Yuta~Hamada,\thanks{E-mail: \tt hamada@gauge.scphys.kyoto-u.ac.jp}~
Hikaru~Kawai,\thanks{E-mail: \tt hkawai@gauge.scphys.kyoto-u.ac.jp}~ 
and Kin-ya~Oda\thanks{E-mail: \tt odakin@phys.sci.osaka-u.ac.jp}\bigskip\\
\it \normalsize
${}^{*\dagger}$Department of Physics, Kyoto University, Kyoto 606-8502, Japan\smallskip\\
\it \normalsize
${}^\ddag$Department of Physics, Osaka University, Osaka 560-0043, Japan
}



\maketitle

\abstract{\noindent \normalsize
We consider a possibility that the Higgs field in the Standard Model (SM) serves as an inflaton when its value is around the Planck scale. We assume that the SM is valid up to an ultraviolet cutoff scale $\Lambda$, which is slightly below the Planck scale, and that the Higgs potential becomes almost flat above $\Lambda$. Contrary to the ordinary Higgs inflation scenario, we do not assume the huge non-minimal coupling, of $O(10^4)$, of the Higgs field to the Ricci scalar. 
We find that $\Lambda$ must be less than $5\times10^{17}\GeV$ in order to explain the observed fluctuation of the cosmic microwave background, no matter how we extrapolate the Higgs potential above $\Lambda$. The scale $10^{17}\GeV$ coincides with the perturbative string scale, which suggests that the SM is directly connected with the string theory. For this to be true, the top quark mass is restricted to around 171\GeV, with which $\Lambda$ can exceed $10^{17}\GeV$. 
As a concrete example of the potential above $\Lambda$, we propose a simple log type potential. The predictions of this specific model for the $e$-foldings $N_*=50$--60 are consistent with the current observation, namely, the scalar spectral index is $n_s=0.977\text{--}0.983$ and the tensor to scalar ratio $0<r<0.012$--0.010. Other parameters, $\df n_s/\df\ln k$, $n_t$, and their derivatives, are also consistent.
}

\newpage

\normalsize

\section{Introduction}
It is more and more plausible that the particle discovered at the CERN Large Hadron Collider (LHC)~\cite{Aad:2012tfa,Chatrchyan:2012ufa} around 126\,GeV is the Standard Model (SM) Higgs boson. Its couplings to the $W$ and $Z$ gauge bosons, to the top and bottom quarks, and to the tau lepton are all consistent to those in the SM within one standard deviation even though their values vary two orders of magnitude, see e.g.\ Ref.~\cite{Giardino:2013bma}.
No hint of new physics beyond the SM has been found so far at the LHC up to 1\,TeV.
It is important to examine up to what scale the SM can be a valid effective description of nature.

The determination of the Higgs mass finally fixes all the parameters in the SM. We can now obtain the \emph{bare} parameters at $\Lambda$. These parameters are important. If a ultraviolet (UV) theory such as string theory fails to fit them, it is killed.

The parameters in the SM are dimensionless except for the Higgs mass (or equivalently its vacuum expectation value (VEV)). The dimensionless bare coupling constants can be approximated by the running ones at $\Lambda$, see e.g.\ Appendix of Ref.~\cite{Hamada:2012bp}. Once the low energy inputs are given, we can evaluate the running couplings through the renormalization group equations (RGEs) of the SM. 
The detailed RGE study of the SM tells us that both the Higgs quartic coupling and its beta function become tiny at the same scale $\sim 10^{17}\GeV$ for the input value of the Higgs mass around 126\GeV; see e.g.\ Refs.~\cite{Holthausen:2011aa,Bezrukov:2012sa,Degrassi:2012ry,Alekhin:2012py,Masina:2012tz,Hamada:2012bp,Jegerlehner:2013cta,Buttazzo:2013uya} for latest analyses.

After fixing all the dimensionless bare couplings, the last remaining parameter in the SM is the bare Higgs mass.
The quadratically divergent bare Higgs mass is found to be suppressed too when the UV cutoff is $\Lambda\gtrsim 10^{17}\GeV$~\cite{Hamada:2012bp}; see also Refs.~\cite{Jegerlehner:2013cta,Jegerlehner:2013nna,Sola:2013gha,Masina:2013wja}, \rev{and also Refs.~\cite{Alsarhi:1991ji,Einhorn:1992um,Kolda:2000wi,Casas:2004gh,Jones:2013aua}}.
The absence of the bare mass at $\Lambda$, along with the vanishing quartic coupling and its beta function, implies that the Higgs potential is approximately flat there and that its height is suppressed compared to (the fourth power of) the cutoff scale.

Following the evidence of the top quark with mass $174\pm10^{+13}_{-12}\GeV$~\cite{Abe:1994xt} in 1994, Froggatt and Nielsen have predicted~\cite{Froggatt:1995rt} that the top and Higgs masses are $173\pm5\,\text{GeV}$ and $135\pm9\,\text{GeV}$, respectively. This prediction is based on the multiple point principle (MPP) that the SM Higgs potential must have another minimum at the Planck scale and that its height is (order-of-magnitude-wise~\cite{Froggatt:2001pa,Nielsen:2012pu}) degenerate to the SM one. This assumption is equivalent to the vanishing Higgs quartic coupling and its beta function at the Planck scale. The success of this prediction indicates that at least the top-Higgs sector of the SM remains unaltered up to a very high UV cutoff scale $\Lambda$.\footnote{ As a possible modification, the classically conformal $B-L$ model is considered in~\cite{Iso:2009ss,Iso:2009nw,Holthausen:2009uc}. This model can realize the flat potential at the Planck scale and solve the hierarchy problem~\cite{Iso:2012jn}.}

In the MPP, it is assumed that there are two vacua that are separated by a potential barrier, as illustrated in the lowermost (green) solid and dashed lines in Fig.~\ref{flat potential}.
In this paper, we consider the case where there is no potential barrier, as illustrated in the middle (blue) and uppermost (red) lines in Fig.~\ref{flat potential} so that the Higgs potential can be used for an inflation. 
In order to have an inflation consistent with the current observational data, we assume that the low energy SM Higgs potential, depicted by the solid lines, is smoothly connected to an almost flat potential, depicted by the dot-dashed lines, around the UV cutoff scale $\Lambda$.\footnote{
In the original argument of the MPP~\cite{Froggatt:1995rt}, the partition function has been maximized as a function of the bare Higgs mass. In more recent Ref.~\cite{Nielsen:2012pu}, Nielsen has generalized their argument and considered several possibilities of the function to be maximized.
In the context of Ref.~\cite{Nielsen:2012pu}, our approach could be regarded as the maximization of the entropy of the universe by requiring the occurrence of the inflation.
}
 
Some of the concrete examples of the flat potential above the cutoff are the following:
In the gauge-Higgs unification scenario, the potential for $\left\langle A_5\right\rangle$ is almost flat for large field values $\left\langle A_5\right\rangle\gg R^{-1}$ because of the gauge invariance and is bounded from above by $O(R^{-4})$~\cite{Hosotani:1983xw}; similar mechanism can be expected in string compactification~\cite{Hebecker:2012qp,Hebecker:2013lha}; see also Refs.~\cite{Haba:2005kc,Haba:2008dr,Maru:2013ooa,Maru:2013bja,Maru:2013qla}.
Another example is the Coleman-Weinberg potential~\cite{Coleman:1973jx} with an explicit momentum cutoff $\Lambda$.
For a field value beyond $\Lambda$, we get a log potential, as we explain in Appendix~\ref{log motivation}. (This possibility is pursued in Section~\ref{log section}.)

The latest cosmological data~\cite{Ade:2013uln} constrains the scalar fluctuation amplitude $A_s$, the spectral index $n_s$, the tensor-to-scalar ratio $r$; see also Ref.~\cite{Martin:2013tda} for a recent review. By assuming the slow-roll inflation, these values are completely fixed by the potential height $\V_*$ and its derivatives $\V_*'$ and $\V_*''$ at a field value $\varphi_*$ corresponding to a given $e$-folding $N_*\simeq60$.
If we allow arbitrary shape of the Higgs potential in all the range beyond the electroweak scale, we can trivially satisfy these constraints.
In our case, however, the SM potential in the range $\varphi<\Lambda$ is fixed by the known SM parameters.
In order to avoid the graceful exit problem~\cite{Linde:1981mu,Hawking:1982ga,Guth:1982pn}, the potential must be monotonically increasing in the entire region, both below and above $\Lambda$. In particular, the value of the SM potential at the cutoff must satisfy $\V_\text{SM}(\varphi=\Lambda)<\V_*$.
As a result, we get an upper bound: $\Lambda\lesssim10^{17}\GeV$, when the top quark mass is $M_t\lesssim 171\GeV$, irrespectively of the extrapolation of the potential above the cutoff $\varphi>\Lambda$.

As said above, we propose a possible extrapolation of the potential beyond $\Lambda$, log plus constant, motivated by the Coleman-Weinberg potential with the cutoff $\Lambda$. We show predictions of the spectral indices, their derivatives, etc.\ of the cosmic microwave background (CMB) in this model.

Our scenario differs from the conventional Higgs inflation scenario~\cite{Bezrukov:2007ep,Bezrukov:2009db,Bezrukov:2010jz,Salvio:2013rja, CervantesCota:1994zf,CervantesCota:1995tz}, which achieves the flatness of the Higgs potential by the huge Higgs coupling to the Ricci scalar: $\xi\ab{\phi}^2\mathcal R$ with $\xi\sim 10^4$. The idea of the Higgs inflation is attractive, but it would be even better if we can realize it without such a coupling to gravity. See also Ref.~\cite{Giudice:2010ka,Lee:2013nv} for the unitarity issue of the conventional Higgs inflation, which necessitates new particles above $M_P/\xi$.\footnote{
The authors of Refs.~\cite{Kamada:2010qe,Kamada:2012se} have proposed a Higgs inflation model in which the Higgs kinetic term is modified; the unitarity issue of this scenario, which involves higher derivatives of the Higgs field, would also be interesting to study.
}

This paper is organized as follows.
In Section~\ref{constraints section}, we review the constraints from the cosmological observations including the latest results from the Planck experiment.
In Section~\ref{SM Higgs potential}, we present the RGE running of the dimensionless couplings in the SM in the modified minimal subtraction ($\overline{\text{MS}}$) scheme; we also show the bare Higgs mass-squared parameter at $\Lambda$. Then we show that the SM Higgs potential can become flat around $10^{17}\GeV$.
In Section~\ref{minimal Higgs inflation}, we examine a necessary condition that the Higgs field in the SM can serve as an inflaton assuming arbitrary shape of potential above $\Lambda$.
In the last section, we summarize our results.

\section{Constraints on inflation models}\label{constraints section}
We briefly review and summarize our notation on the cosmological constraints from the CMB data observed at the Planck experiment, basically following Ref.~\cite{Ade:2013uln}.
The curvature and tensor power spectra are expanded around a pivot scale $k_*$ as
\al{
\mc P_{\mc R}
	&=	A_s\,\paren{k\over k_*}^{
			n_s-1
			+{1\over2}{\df n_s\over\df\ln k}\ln{k\over k_*}
			+{1\over3!}{\df^2 n_s\over{\df\ln k}^2}\paren{\ln{k\over k_*}}^2
			+\cdots
			},	&
\mc P_t
	&=	A_t\,\paren{k\over k_*}^{
			n_t
			+{1\over2}{\df n_t\over\df\ln k}\ln{k\over k_*}
			+\cdots
			}.
}
We take the slow-roll approximation hereafter. 
The slow-roll parameters at a given position $\varphi$ of the inflaton potential $\V$ are defined as
\al{
\epsilon_V
	&=	{M_P^2\over2}{\V_\varphi^2\over \V^2},	&
\eta_V
	&=	M_P^2{\V_{\varphi\varphi}\over \V},&
\xi_V^2
	&=	M_P^4{\V_\varphi \V_{\varphi\varphi\varphi}\over \V^2},	&
\varpi_V^3
	&=	M_P^6{\V_\varphi^2\,\V_{\varphi\varphi\varphi\varphi}\over \V^3}.
}
The number of $e$-folding before the end of inflation $t_\text{end}$ from a time $t_*$ becomes
\al{
N_*	&=	\int_{t_*}^{t_\text{end}}\df t\,H
	=	\int_{\varphi_*}^{\varphi_\text{end}}{\df\varphi\over\dot\varphi}H
	=
		{1\over M_P^2}\int_{\varphi_\text{end}}^{\varphi_*}{\V\over \V_\varphi}\df\varphi
	=	\frac{1}{M_P}\int_{\varphi_\text{end}}^{\varphi_*} \frac{\df\varphi}{\sqrt{2\epsilon_V}},
}
where we have taken $\V_\varphi>0$ in the last step.
The end of inflation is defined by the field value $\varphi_\text{end}$ below which the slow roll condition is violated:
\al{
\max\Big\{\epsilon_V(\varphi_\text{end}),\,\big|\eta_V(\varphi_\text{end})\big|\Big\}=1.
	\label{end of inflation}
}
For most reasonable inflation models, the scale that we are observing from the CMB data corresponds to the $e$-folding in the range~\cite{Ade:2013uln}
\al{
50<N_*<60.
	\label{N range}
}
In the following, we evaluate the slow-roll parameters at $\varphi_*$ that satisfies Eq.~\eqref{N range}. (We will also consider the range $40<N_*<50$ in Section \ref{log section} for comparison.)

The cosmological parameters are given by\footnote{
\rev{
\Magenta{Eq.~(17) of Ref.~\cite{Ade:2013uln} contains a typo (an overall wrong sign for $\df n_s/\df\ln k$). Also note that in the case of $\Lambda\text{CDM}+r+\df n_s/\df\ln k$ in Table 5 of Ref.~\cite{Ade:2013uln}, there is missing minus in front of the mean value of the $\df n_s/\df\ln k$ for Planck+WP and Planck+WP+lensing.}
We thank \addedSecond{the referee} \Magenta{for bringing this point to our attention and Finelli Fabio for his kind clarification}.
	}
}
\al{
A_s	&=
		{\V\over24\pi^2M_P^4\epsilon_V},	&
A_t	&=
		{2\V\over 3\pi^2M_P^4},	&
r	&=	{A_t\over A_s}
	=
		16\epsilon_V,
		\label{overall normalization}
}
\al{
n_s	&=
		1+2\eta_V-6\epsilon_V,	&
n_t	&=
		-2\epsilon_V,	\nn
{\df n_s\over\df\ln k}
	&=
		\Magenta{16\epsilon_V\eta_V-24\epsilon_V^2-2\xi_V^2},	&
{\df n_t\over\df\ln k}
	&=
		-4\epsilon_V\eta_V+8\epsilon_V^2,	\nn
{\df^2 n_s\over{\df\ln k}^2}
	&=
		-192\epsilon_V^3+192\epsilon_V^2\eta_V-32\epsilon_V\eta_V^2\nn
	&\quad
		-24\epsilon_V\xi_V^2+2\eta_V\xi_V^2+2\varpi_V^3,
}
where the quantities are evaluated at the field value $\varphi_*$.

At the pivot scale $k_*=0.05\,\text{Mpc}^{-1}$, the scalar amplitude $A_s$ and the spectral index~$n_s$ are constrained by the Planck+WMAP data as~\cite{Ade:2013uln}
\al{
A_s
	&=	\paren{2.196^{+0.051}_{-0.060}}\times10^{-9},
		\label{As bound}
}
\al{
n_s	&=	0.9603\pm0.0073,
	\label{single parameter constraint}
}
assuming ${\df n_s/\df\ln k}={\df^2 n_s/{\df\ln k}^2}=r=0$.

If we include the tensor-to-scalar ratio $r$ as an extra parameter, 
the Planck+WMAP+high-$\ell$ data~\cite{Ade:2013uln} at the pivot scale $k_*=0.002\,\text{Mpc}^{-1}$ give the 1$\sigma$ range for $n_s$ and the 95\% CL limit on $r$ as
\al{
n_s	&=	0.9600\pm 0.0071,	&
r	&<	0.11.	\label{tensor to scalar ratio}
}

On the other hand, if we include ${\df n_s/\df\ln k}$ as an extra parameter, we obtain the constraint at the pivot scale $k_*=0.05\,\text{Mpc}^{-1}$~\cite{Ade:2013uln}
\al{
n_s	&=	0.9561\pm0.0080,	&
{\df n_s\over\df\ln k}
	&=	-0.0134\pm0.0090.
		\label{running index constraint}
}



%
One may vary both $r$ and $\df n_s/\df \ln k$ to fit the Planck+WMAP data at the pivot scale $k_*=0.05\,\text{Mpc}^{-1}$, and obtains~\cite{Ade:2013uln}
\al{
n_s&=0.9583\pm0.0081,   &
r&<0.25,  &
\frac{\df n_s}{\df \ln k}&=\cmag{-}0.021\pm0.012,
}
where the constraint on $r$ is given at 95\% CL.
\footnote{
In terms of the slow-roll parameters, these conditions become
\als{
\epsilon_V	&<	0.015, &
\eta_V
	&=	-0.014^{+0.015}_{-0.011},	&
\ab{\xi_V^2}
	&=	0.009\pm0.006.
}
where the constraint on $\epsilon_V$ is given at 95\% CL.
}

We note that the upper bound on $r$ gives that of the inflaton energy scale~\cite{Ade:2013uln}
\al{
\V_{\text{max}}
	&=	{3\pi^2A_s\over2}r_{\text{max}}M_P^4
	=	1.3\times10^{65}\text{GeV}^4\times\paren{r_{\text{max}}\over0.11}.
		\label{potential condition}
}

\section{SM Higgs potential}\label{SM Higgs potential}
The SM Higgs potential much above the electroweak scale but below the cutoff scale $\Lambda$ is governed by the RGE running of the Higgs quartic coupling $\lambda$, which highly depends on the top Yukawa coupling~$y_t$.
Therefore we first review how $y_t$ is determined;
then we show the numerical results of the RGEs;
finally we present the resultant Higgs potential around $\Lambda$.

\subsection{Coupling constants at the electroweak scale}
The most precise determination of the top quark mass is given by a combination of the Tevatron data for the invariant mass of the top quark decay products~\cite{CDF:2013jga}:
\al{
M_t^\text{inv}=173.20\pm0.87\GeV.
	\label{Tevatron mass}
}
The problem of the Tevatron determination~\eqref{Tevatron mass} is that the invariant mass, which is reconstructed from the color singlet final states, cannot be the pole mass of the colored top quark~\cite{Alekhin:2012py}. Instead, the authors of Ref.~\cite{Alekhin:2012py} proposed to get the top mass by fitting the $t\bar t+X$ inclusive cross section, and obtained the pole mass:\footnote{
As clarified in Refs.~\cite{Hamada:2013cta,Jegerlehner:2013dpa}, currently there are two ways to define the \MSbar running top mass for a given \MSbar running Yukawa $y_t(\mu)$.
The \MSbar mass used in QCD~\cite{Abazov:2011pta,Aldaya:2012tv,Alekhin:2012py}, which we call $m_t^\text{QCD}(\mu)$, can be approximately written as $m_t^\text{QCD}(\mu)\simeq y_t(\mu)V/\sqrt{2}$ with $V=246.22\GeV$, up to electroweak corrections less than $1\%$. In Refs.~\cite{Hempfling:1994ar,Jegerlehner:2012kn}, \MSbar mass is defined as $m_t(\mu):=y_t(\mu)v(\mu)/\sqrt{2}$, where $v(\mu)$ is given by the relation $-m^2(\mu)=\lambda(\mu)\,v^2(\mu)$, with $m^2(\mu)$ being the running mass parameter in the tree potential in the \MSbar scheme: $\V=m^2(\mu)\,\phi^\dagger\phi+\lambda(\mu)\,(\phi^\dagger\phi)^2$. There are $\sim7\%$ difference between $m_t^\text{QCD}(M_t)$ and $m_t(M_t)$~\cite{Jegerlehner:2012kn}, which is mainly due to the tadpole contribution from the top quark.
Though the bound on the pole mass~\eqref{Djouadi mass} has been derived from that on $m_t^\text{QCD}(M_t)=163.3\pm2.7\GeV$~\cite{Alekhin:2012py}, it is consistent to use Eq.~\eqref{Djouadi mass} in obtaining the Yukawa coupling~\eqref{Strumia Yukawa} since the pole mass $M_t$ should be the same in both schemes. 
}
\al{
M_t	&=	173.3\pm2.8\GeV.	\label{Djouadi mass}
}
The numerical value of the \MSbar Yukawa coupling at the top mass scale can be read off from Ref.~\cite{Degrassi:2012ry} as
\al{
y_t(M_t)
	&=	0.93669
		+0.01560\left(M_t-173.3\GeV\over2.8\GeV\right)
		-0.00041\left(\alpha_s(M_Z)-0.1184\over0.0007\right)\nn
	&\quad
		-0.00001\left({M_H-125.6\GeV\over0.4\GeV}\right)
		\pm0.00200_\text{th},
		\label{Strumia Yukawa}
}
where we have employed a combined Higgs mass $M_H=125.6\pm0.4\GeV$.
The electroweak gauge couplings at the $Z$ mass scale are~\cite{Beringer:1900zz}
\al{
g_Y(M_Z)&=0.357418(35),	&
g_2(M_Z)&=0.65184(18).
}
The \MSbar strong and quartic couplings at the top mass scale are~\cite{Degrassi:2012ry}
\al{\label{strong coupling}
g_s(M_t)
	&=	1.1644+0.0031\left(\frac{\alpha_s(M_Z)-0.1184}{0.0007}\right)-0.0013\left({M_t-173.3\GeV\over2.8\GeV}\right),\\
\lambda(M_t)
	&=	0.12699+0.00082\left({M_H-125.6\GeV\over0.4\GeV}\right)-0.00012\left({M_t-173.3\GeV\over2.8\GeV}\right)\pm 0.00140_{\text{th}}.
}

\subsection{Numerical Results of SM RGEs}
\begin{figure}[tn]
\begin{center}
\includegraphics[width=.49\textwidth]{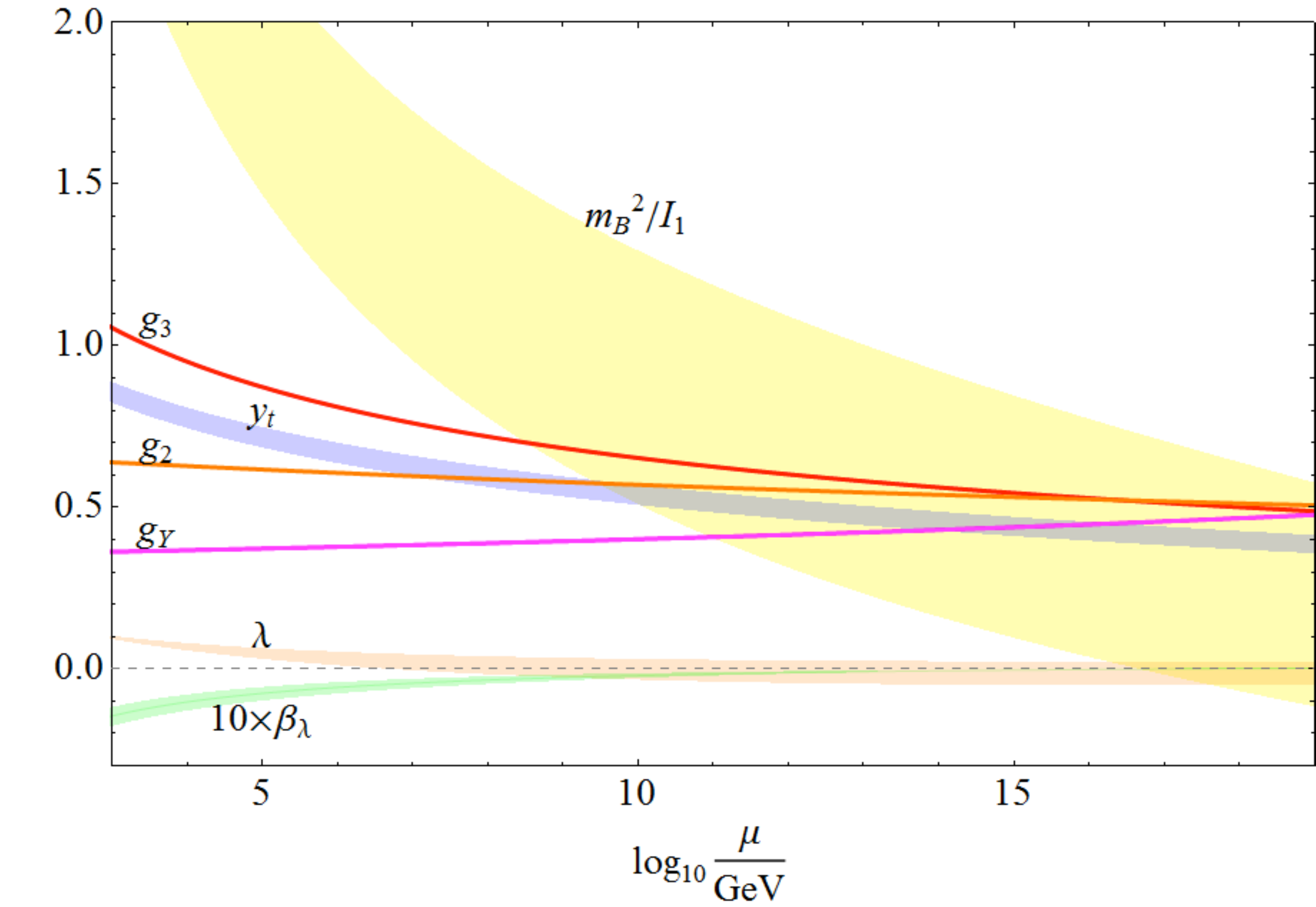}
\hfill
\includegraphics[width=.49\textwidth]{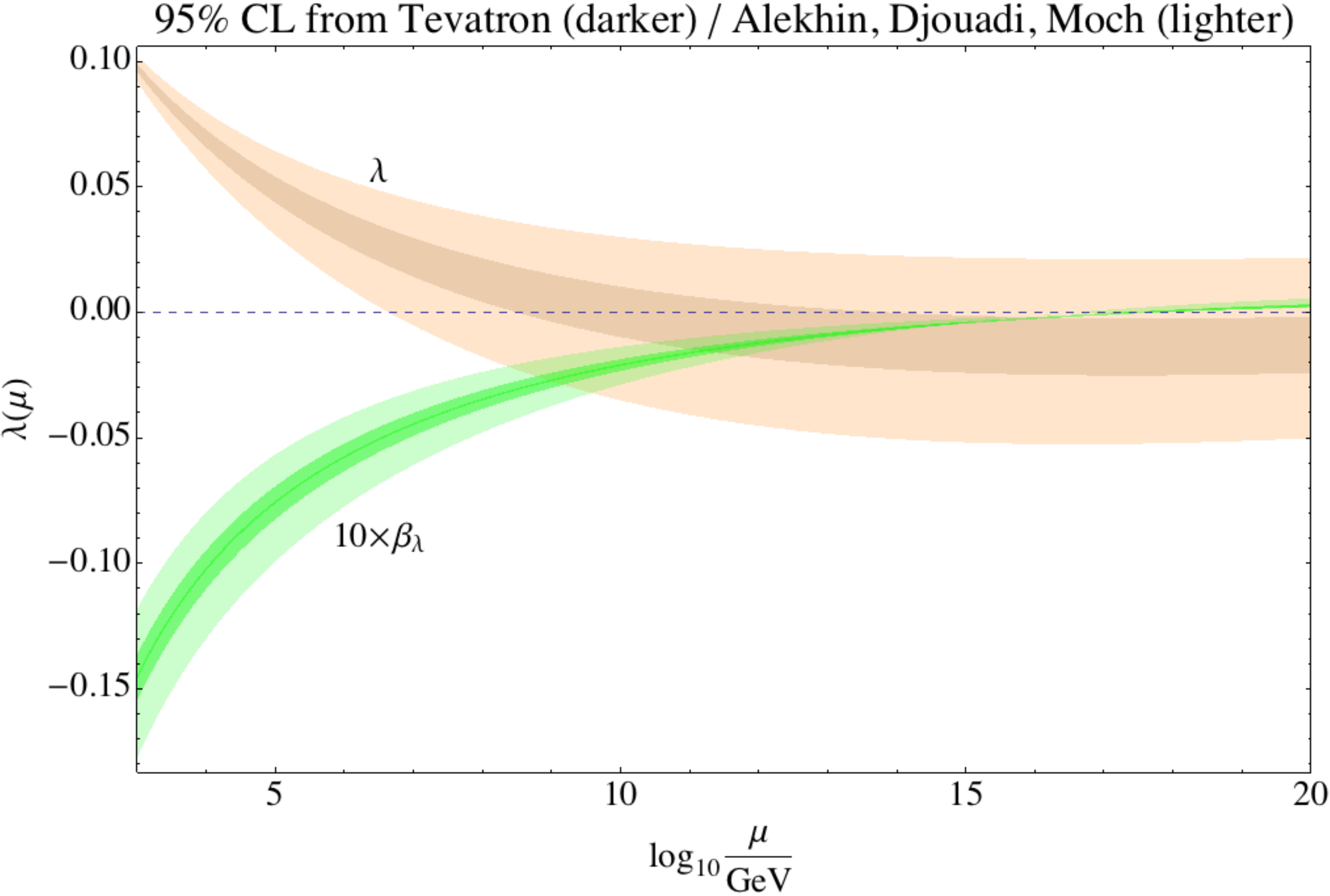}
\caption{Left: \MSbar running couplings. 
The 95\% confidence intervals are given for $m_B^2/I_1$, $y_t(\mu)$, $\lambda(\mu)$, and $10\beta_\lambda(\mu)$; see text for more details. The intervals for the gauge couplings $g_Y$, $g_2$, and $g_3$ are too small to be seen.
Right: Enlarged view around the horizontal axis of Left.
Darker bands are the 95\% confidence intervals under the (theoretically unjustified) assumption that the Tevatron mass~\eqref{Tevatron mass} can be identified with the top pole mass $M_t$. }\label{SMcouplings}
\end{center}
\end{figure}
We use the two-loop RGEs in the SM which are summarized in Ref.~\cite{Hamada:2012bp}.
We show our result of running \MSbar couplings in Fig.~\ref{SMcouplings}.
The gauge couplings $g_Y$, $g_2$, $g_3$ are drawn by thick lines.
The thickness of the curves for $y_t, \lambda$ and $\beta_\lambda$ comes from the 1.96$\sigma$ variation of $M_t$~\eqref{Djouadi mass}, where $\alpha_s(M_z)$ and $M_H$ are fixed to their central values.
Similarly, we plot the bare Higgs mass-squared $m_B^2$, divided by the quadratically divergent integral $I_1=\Lambda^2/16\pi^2$, as a function of $\Lambda$~\cite{Hamada:2012bp,Hamada:2013cta}. Note that the bare mass $m_B^2$ is \emph{not} the running mass.

We see that the Higgs quartic coupling $\lambda$ has a minimum around $10^{17}\GeV$. 
This is due to the fact that the beta function of $\lambda$ receives less negative contribution from the top loop since $y_t$ becomes smaller at high scales.

We can fit the parameters
at the reduced Planck scale\footnote{
In this definition, the graviton fluctuation $h_{\mu\nu}$ around the flat Minkowski spacetime: $g_{\mu\nu}=\eta_{\mu\nu}+2M_P^{-1}h_{\mu\nu}$ becomes canonically normalized.
}
$M_P:=1/\sqrt{8\pi G}=2.4\times10^{18}\GeV$
as
\al{
\lambda(M_P)
	&=	-0.015
		-0.019\paren{{M_t-173.3\GeV\over2.8\GeV}}\nn
	&\quad
		+0.002\paren{\alpha_s(M_Z)-0.1184\over0.0007}
		+0.001\paren{{M_H-125.6\GeV\over0.4\GeV}},	\\
{m_B^2\over M_P^2/16\pi^2}
	&=	0.26
		+0.18\paren{{M_t-173.3\GeV\over2.8\GeV}}
		-0.02\paren{\alpha_s(M_Z)-0.1184\over0.0007}
		-0.01\paren{{M_H-125.6\GeV\over0.4\GeV}},	\\
\beta_\lambda
	&=	0.000103
		+0.000069\paren{{M_t-173.3\GeV\over2.8\GeV}}
		+0.000028\paren{{M_t-173.3\GeV\over2.8\GeV}}^2\nn
	&\quad
		-0.000013\paren{{M_H-125.6\GeV\over0.4\GeV}},
}
where the dependence of $\beta_\lambda$ on $\alpha_s(M_Z)$ is of $O(10^{-7})$ and is not shown.


\subsection{Higgs inflation?}\label{Higgs potential in SM}
\begin{figure}[tn]
\begin{center}
\hfill
\includegraphics[width=.6\textwidth]{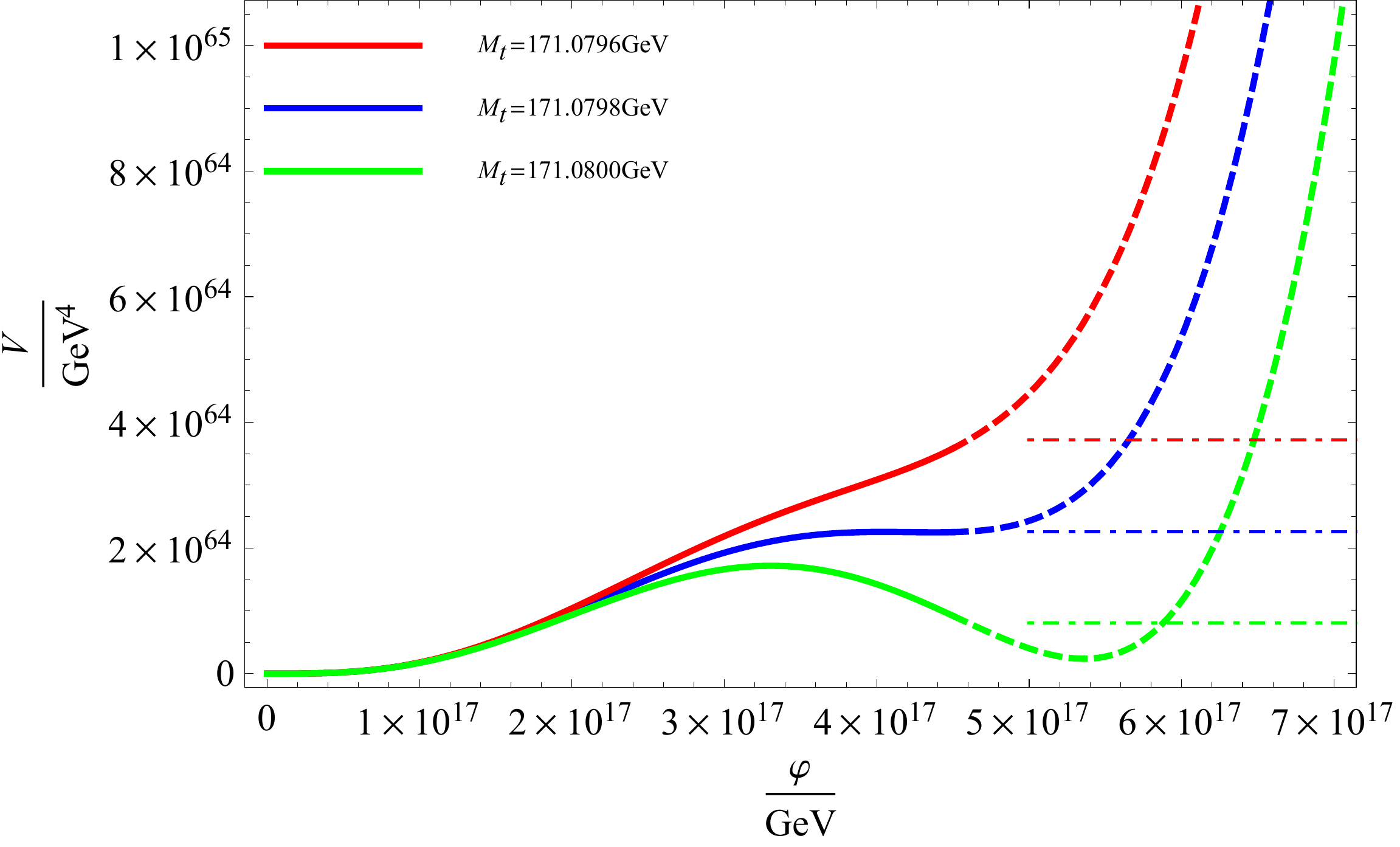}
\hfill\mbox{}
\caption{RGE improved Higgs potential~\eqref{Higgs potential}, with $M_H=125.6\GeV$ and $\alpha_s=0.1184$.
The Solid and dashed lines are the SM Higgs potential~\eqref{Higgs potential}. Beyond the UV cutoff $\Lambda$, which we have taken \rev{in this figure} to be $4.5\times 10^{17}\GeV$ as an illustration, we assume that the potential becomes flat as depicted by the dot-dashed lines.
}\label{flat potential}
\end{center}
\end{figure}

We have seen in Fig.~\ref{SMcouplings} that both the Higgs self coupling $\lambda$ and its beta function~$\beta_\lambda$ become very small at high scales $\mu\gtrsim 10^{17}\GeV$.
This fact suggests that the SM Higgs field could be identified as an inflaton.
In the following, we show that the SM Higgs potential becomes flat 
around $10^{17}\GeV$ if we tune the top quark mass. However, we will see that it is difficult to reconcile this potential with the cosmological observation~\cred{\cite{Isidori:2007vm,Nielsen:2012pu}}.

For a field value $V=246.22\GeV\ll\varphi<\Lambda $, the Higgs potential becomes, with RGE improvement,
\al{
\V_\text{SM}(\varphi)
	&=	\frac{\lambda(\varphi)}{4}\varphi^4.
		\label{Higgs potential}
}
Around the scale $10^{17}\GeV$, this potential strongly depends on the top quark mass, which we show by the solid and dashed lines in Fig.~\ref{flat potential}. The dot-dashed lines are irrelevant to the argument of this subsection.

If we fine tune the top mass, we can have a saddle point in the Higgs potential: $\V_\varphi(\varphi)=\V_{\varphi\varphi}(\varphi)=0$, as indicated by the middle (blue) line in Fig.~\ref{flat potential}. When we slightly lower the top mass, the saddle point disappears and the potential becomes monotonically increasing, as the upper (red) line. On the contrary, when we slightly raise the top mass, there appears another minimum at a high scale, as the lower (green) line~\cite{Isidori:2001bm,Branchina:2013jra}.

The middle (blue) line case, $M_t=171.0798\GeV$, gives the potential
\al{
\V_c
	&=		6.0\times10^{-10}M_P^4
	\sim	\paren{10^{16}\GeV}^4
			\label{potential height}
}
at the saddle point $\varphi_c=4.2\times10^{17}\GeV$.
One might think of using this saddle point for a Higgs inflation, but it is impossible due to the following reasons:
With Eq.~\eqref{As bound}, this height of potential necessitates $\epsilon_V\sim10^{-3}$.
However, the point of $N_*\gtrsim 50$ becomes too close to the saddle point and gives $\epsilon_V\ll10^{-3}$. 

In order to avoid this problem, one might try lowering the top mass slightly to reproduce the value of $\epsilon_V\sim10^{-3}$ at the inflection point $\V_{\varphi\varphi}=0$. This still does not work because we cannot have enough $e$-foldings, $N_*\gtrsim50$, at the inflection point. 

As the third trial, one might choose $\epsilon_V$ freely at the inflection point so that one can have enough $e$-folding in passing the inflection point. In this case, one tries to reproduce $\epsilon_V\sim10^{-3}$ at the higher point with $N_*\sim 50$. However, $\eta_V$ at this point turns out to be too large to satisfy the slow-roll condition. 

We present more detailed discussion in Appendix~\ref{SM Higgs as inflaton}.


Note that the precise value of $M_t$ to give the saddle point in Fig.~\ref{flat potential} depends on $M_H$, the details of the RGEs, etc.; the digits of $M_t$ should not be taken literally, but be regarded as an indication how finely $M_t$ must be tuned to yield a saddle point in the SM. Also the height of the potential at the saddle point varies when we change e.g.\ $M_H$ within the 95\% CL, $M_H=\paren{124.8\,\text{--}\,126.4}\GeV$, as
\al{
\V_c
	&=	\paren{1.5\,\text{--}\,32.}\times10^{-10}M_P^4.
}
Therefore the value~\eqref{potential height} should be taken as an indication of the order of magnitude.

\section{Minimal Higgs inflation}\label{minimal Higgs inflation}
We pursue the possibility that the Higgs potential above $\Lambda$ becomes sufficiently flat to realize a viable inflation, as the dot-dashed lines in Fig.~\ref{flat potential}. The inflaton potential is bounded from above as in Eq.~\eqref{As bound}. In order to avoid the graceful exit problem, the Higgs potential must be monotonically increasing in all the range below and above $\Lambda$. Therefore, even if we allow an arbitrary modification above $\Lambda$,
we \cred{still} can get a bound: $\Lambda<5\times10^{17}\GeV$. As a concrete example of the modification above $\Lambda$, we propose a log type potential and study its cosmological implications.



\subsection{Constraint on top mass from minimal Higgs inflation}

\begin{figure}[tn]
\begin{center}
\hfill
\includegraphics[width=.49\textwidth]{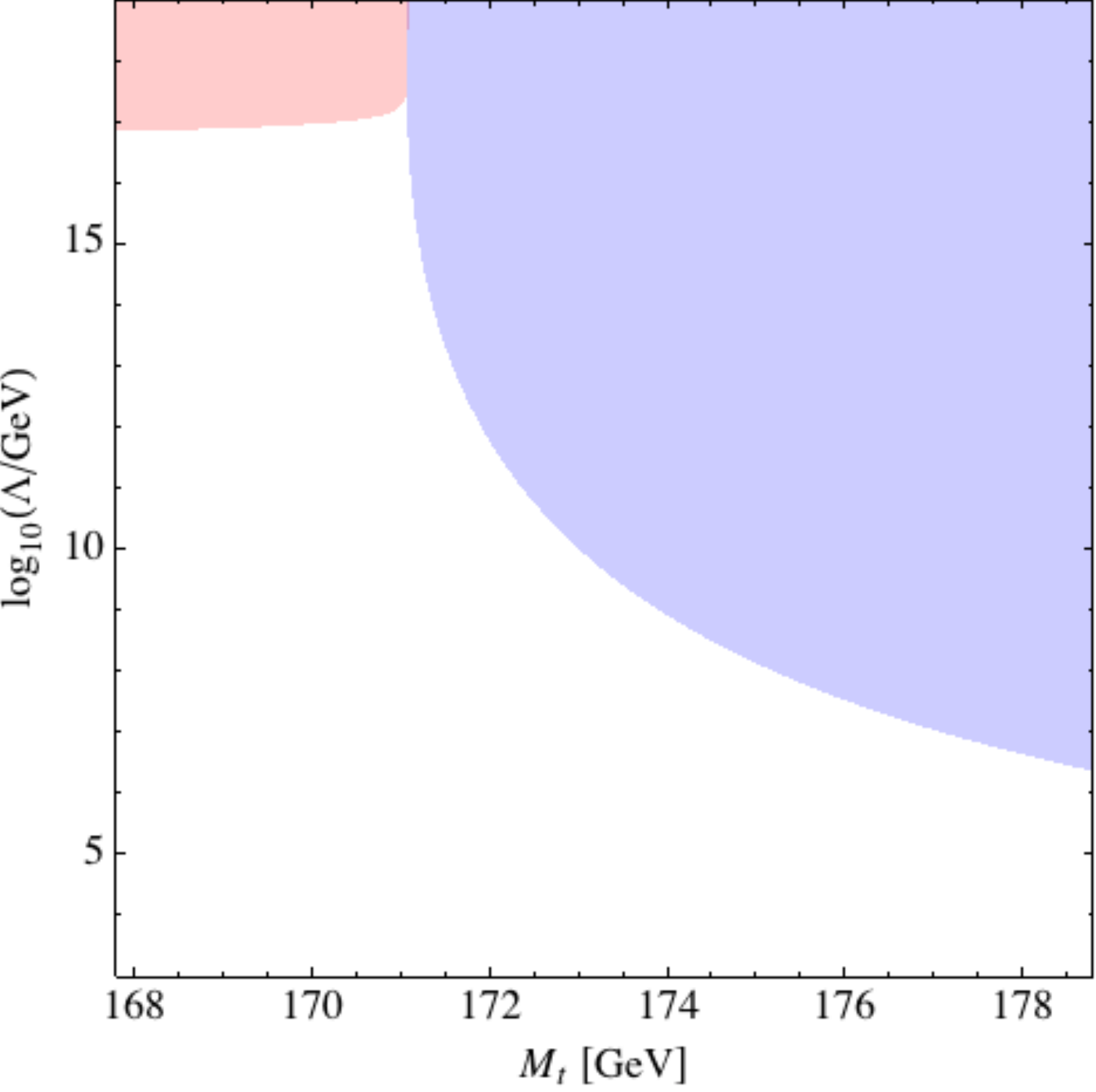}
\hfill
\includegraphics[width=.39\textwidth]{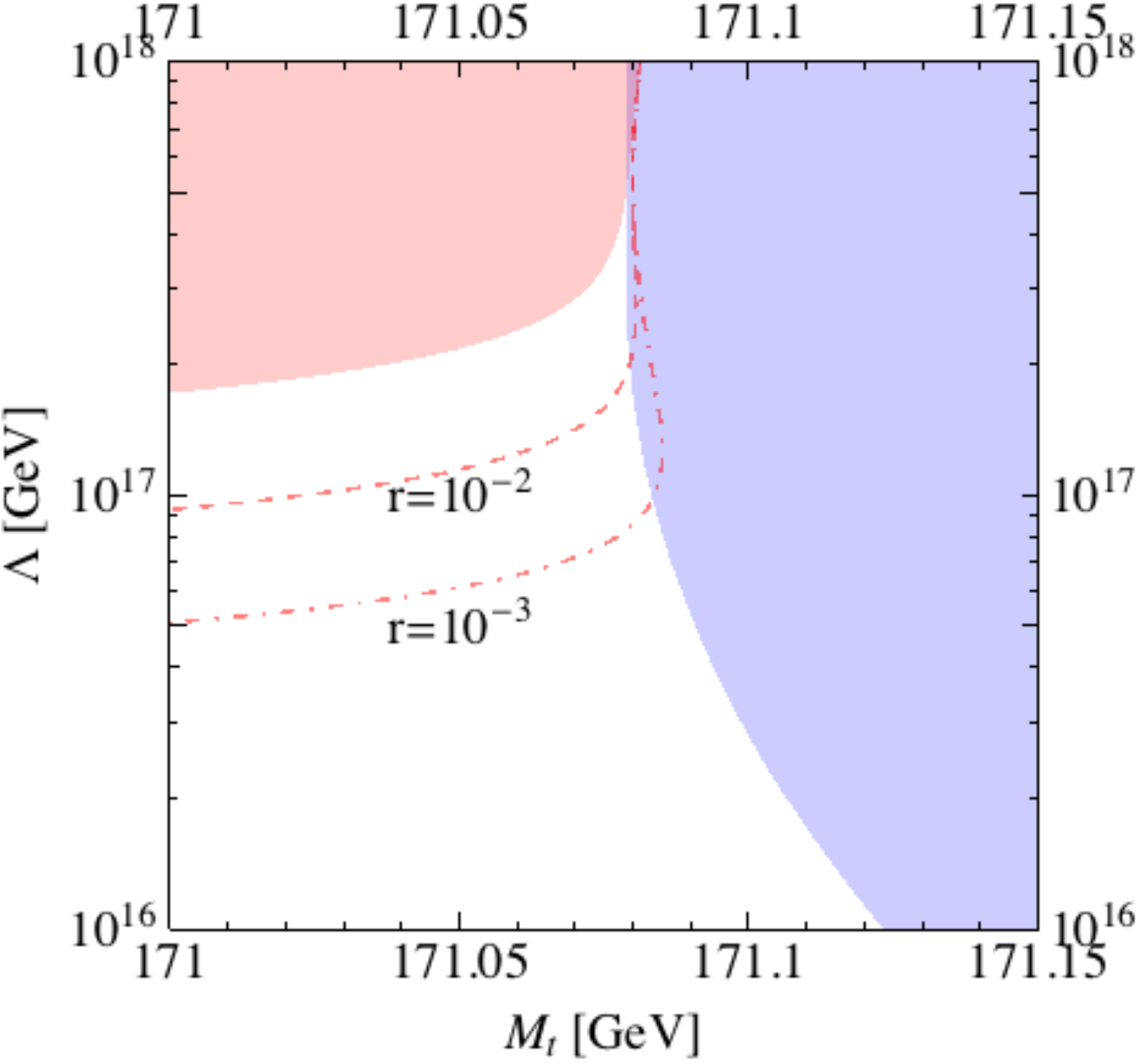}
\hfill\mbox{}
\caption{Left: Excluded region by Eq.~\eqref{SM potential condition} (red, left) and by Eq.~\eqref{derivative condition} (blue, right) in $\log_{10}(\Lambda/\text{GeV})$ vs $M_t$ plane. Right: Enlarged view for $\Lambda$ vs $M_t$. Expected future exclusion limits within 95\%CL: $r<10^{-2}$ and $10^{-3}$ are also presented by dashed and dot-dashed lines, respectively.}\label{topconstraint}
\end{center}
\end{figure}

We have seen that the scale $10^{17}\GeV$ gives the vanishing beta function $\beta_\lambda$. This scale is close to the string scale in the conventional perturbative superstring theory. 
Above the string scale, a conventional local field theory is altered.
We have shown that the bare Higgs mass becomes very small around this scale~\cite{Hamada:2012bp}. This fact strongly suggests that the Higgs boson is a zero mass state of string theory. If it is the case, after integrating out all the massive stringy states, we get the effective potential, which is meaningful for field values beyond the string scale. The resultant potential beyond the string scale would be greatly modified from that in the SM.\footnote{
On the contrary, if the Higgs boson had a string-scale bare mass, it would come from a string massive mode. Then considering the effective potential would become meaningless, as it would become one of the fields to be integrated out.
}

As we have discussed in Introduction, it is plausible that the effective potential for the field value beyond $\Lambda$ becomes almost flat. This opens up a possibility that this flat potential can be used for the inflation. 
Firstly in this subsection, we consider a necessary condition for the SM potential at $\varphi<\Lambda$ to allow such a modification in the region $\varphi>\Lambda$.


To avoid the graceful exit problem~\cite{Linde:1981mu,Hawking:1982ga,Guth:1982pn}, the Higgs potential must be a monotonically increasing function of $\varphi$ in all the range below and above $\Lambda$.\footnote{
If there is a potential barrier, then the phase transition becomes first order. This is problematic: The false vacuum decays only through the tunneling. The bubbles of true vacuum expand with speed of light in the exponentially expanding medium of the false vacuum. They can hardly collide each other.
See also Ref.~\cite{Masina:2012tz} for a possible false vacuum inflation assuming a lowered Planck scale, which we do not employ in this paper.
}
Therefore, we have
\al{
\frac{\df \V_\text{SM}}{\df\varphi}
	&=	\frac{1}{4}\paren{\beta_\lambda+4 \lambda}\varphi^3
	>	0,
\label{derivative condition}
}
for all the scales below the cutoff: $\varphi<\Lambda$, and
the upper bound~\eqref{potential condition} leads to
\al{
\V_\text{SM}(\Lambda)
	&<	\V_\text{max}.
		\label{SM potential condition} 
}

We show excluded regions on the $\Lambda$-$M_t$ plane from the above two constraints in the left panel of Fig.~\ref{topconstraint}. The left (red) region is excluded by the condition~\eqref{SM potential condition} within 95\% CL and the right (blue) region is forbidden by Eq.~\eqref{derivative condition}. The right panel is an enlarged view. The dashed and dot-dashed lines correspond to the exclusion limits at the 95\% CL, $r<10^{-2}$ and $10^{-3}$, that are expected from the future experiments EPIC~\cite{Bock:2009xw} and COrE~\cite{Bouchet:2011ck}, respectively.


We see that the top quark mass needs to be $M_t\lesssim171\GeV$ if we want to have the cutoff scale to be at the string scale $\Lambda\sim10^{17}\GeV$. If the top quark mass turns out to be heavier, say $M_t\gtrsim173\GeV$, then this minimal scenario breaks down. However, it is possible that there exists an extra gauge-singlet scalar $X$ that couples to the SM Higgs boson e.g.\ as
\al{
\mathcal L
	&=	-{\rho\over4!} X^4-{\kappa\over2}\phi^\dagger\phi X^2,
}
where $\rho$ and $\kappa$ are coupling constants. Then $X$ contributes to the running of $\lambda$  positively, and the vacuum stability condition becomes milder. Such a scalar naturally arises in the Higgs portal dark matter scenario; see e.g.\ Refs.~\cite{Silveira:1985rk,McDonald:1993ex,Burgess:2000yq,Davoudiasl:2004be,Patt:2006fw}.\footnote{
No matter $X$ is included or not, our scenario is not altered by the right-handed neutrinos if their Dirac Yukawa couplings are $\lesssim 0.1$.
}



\subsection{Log type potential}\label{log section}
So far we have not specified anything about the potential shape above $\Lambda$. 
In the following, let us examine the log-type potential:
\al{
\V(\varphi)
	&=	\V_1\paren{C+\ln{\varphi\over M_P}}.
		\label{log potential R}
}
We note that the Coleman-Weinberg potential with an explicit momentum cutoff $\Lambda$ leads to a log type potential; see Appendix~\ref{log motivation}.\footnote{
In terms of the parameters given there, $C={\V_0\over\V_1}+\ln{M_P\over\Lambda}$.
}
The potential~\eqref{log potential R} leads to the slow roll parameters
\al{
\epsilon_V
	&=	{1\over2}\paren{M_P\over\varphi}^2\paren{1\over C+\ln{\varphi\over M_P}}^2,	&
\eta_V
	&=	-\paren{M_P\over\varphi}^2{1\over C+\ln{\varphi\over M_P}},
		\nn
\xi_V^2
	&=	2\paren{M_P\over\varphi}^4\paren{1\over C+\ln{\varphi\over M_P}}^2,	&
\varpi_V^3
	&=	-6\paren{M_P\over\varphi}^6\paren{1\over C+\ln{\varphi\over M_P}}^3.
		\label{slow roll parameters}
}

The end point of the inflation is determined from Eq.~\eqref{end of inflation} for a given constant~$C$:
\al{
\varphi_\text{end}
	&=	\begin{cases}
			{M_P\over\sqrt{2}\,W\fn{e^C/\sqrt{2}}} 
				&	\text{for $0<C<0.153$},\\
			\sqrt{2\over W\fn{2e^{2C}}}\,M_P
				&	\text{for $C>0.153$},
		\end{cases}
}
where $W$ is the Lambert function defined by $z=W(z)e^{W(z)}$.
Equivalently, the constant $C$ is fixed as a function of $\varphi_\text{end}$:
\al{
C	&=	\begin{cases}
			{M_P\over\sqrt{2}\varphi_\text{end}}-\ln{\varphi_\text{end}\over M_P}	&	
				\text{for $\varphi_\text{end}\geq\sqrt{2}M_P$,}\\
			\paren{M_P\over \varphi_\text{end}}^2-\ln{\varphi_\text{end}\over M_P}	&	
				\text{for $\varphi_\text{end}\leq\sqrt{2}M_P$.}
		\end{cases}
		\label{R of phi_end}
}
The $e$-folding becomes
\al{
N_*
	&=	{2C-1\over4}{\varphi_*^2-\varphi_\text{end}^2\over M_P^2}
		+{\varphi_*^2\over 2M_P^2}\ln{\varphi_*\over M_P}
		-{\varphi_\text{end}^2\over 2M_P^2}\ln{\varphi_\text{end}\over M_P}\nn
	&=	{\varphi_*^2\over2M_P^2}\ln{\varphi_*\over\varphi_\text{end}}
		-{\varphi_*^2-\varphi_\text{end}^2\over 4M_P^2}
		\times
		\begin{cases}
		\paren{1-\sqrt{2}M_P/\varphi_\text{end}}
			&	\text{for $\varphi_\text{end}\geq\sqrt{2}M_P$},\\
		\paren{1-2M_P^2/\varphi_\text{end}^2}
			&	\text{for $\varphi_\text{end}\leq\sqrt{2}M_P$}.
		\end{cases}
		\label{e-folding for log potential}
}

\begin{figure}[tn]
\begin{center}
\hfill
\includegraphics[width=.7\textwidth]{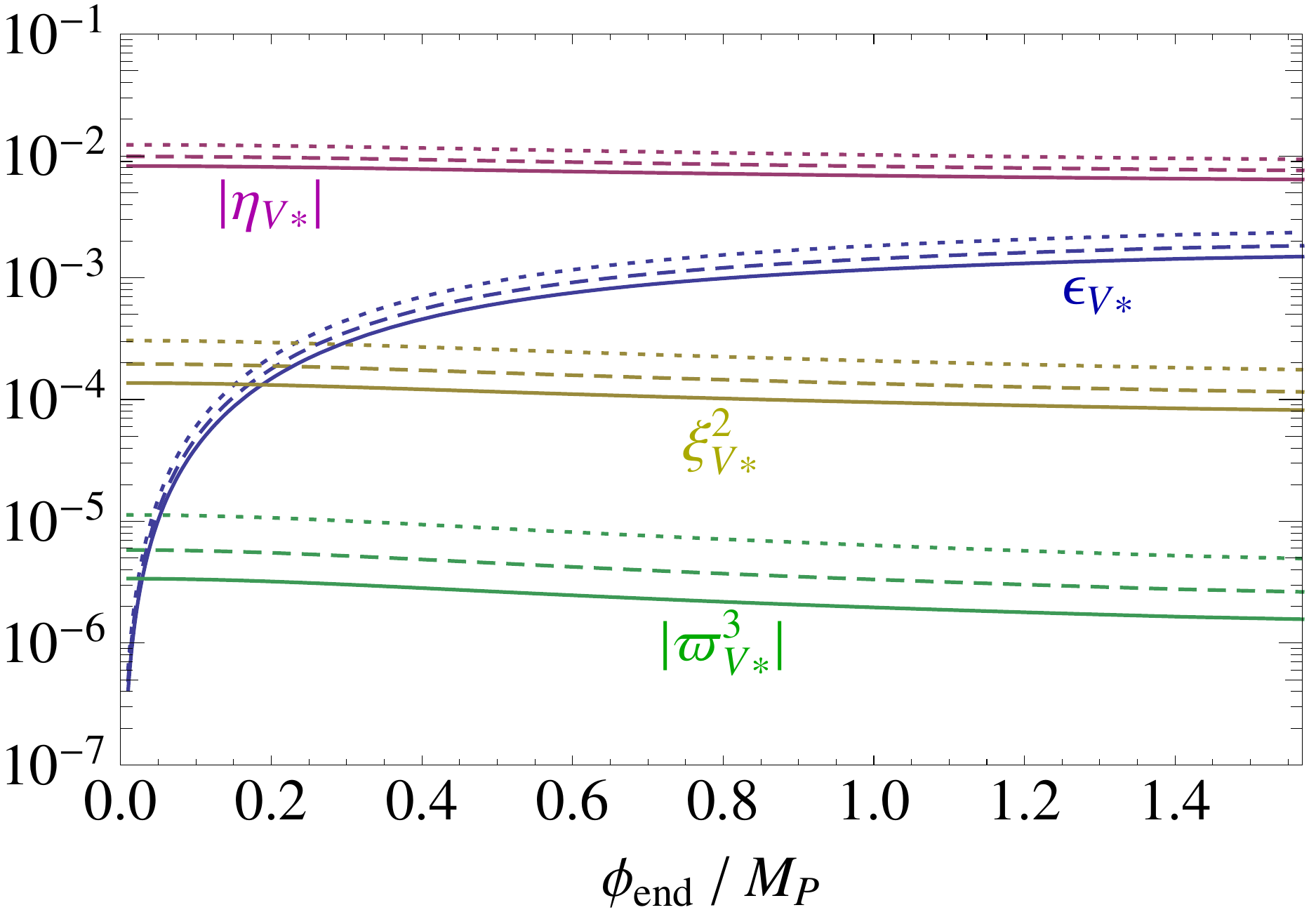}
\hfill\mbox{}
\caption{Slow roll parameters $\epsilon_{V*}$, $\eta_{V*}$, $\xi_{V*}^2$ and $\varpi_{V*}^3$ as functions of $\varphi_\text{end}/M_P$ within the range $0.01<\varphi_\text{end}/M_P<1.57$. The dotted, dashed, and solid lines correspond to $N_*=40$, 50, and 60, respectively.}\label{slow roll vs phi_end}
\end{center}
\end{figure}
To summarize: For a given $\varphi_\text{end}$, we fix the constant $C$ by Eq.~\eqref{R of phi_end}.
Then we can obtain the slow roll parameters from Eq.~\eqref{slow roll parameters} at any field value $\varphi$.
The field value~$\varphi_*$ corresponding to a relevant $e$-folding $N_*$ is determined from Eq.~\eqref{e-folding for log potential}.

This way the slow roll parameters~\eqref{slow roll parameters} at a given $N_*$ is completely fixed.
Note that they are independent of $\V_1$, the overall normalization of the potential.
In Fig.~\ref{slow roll vs phi_end}, we plot the slow roll parameters $\epsilon_V$, $\eta_V$, $\xi_V^2$, and $\varpi_V^3$ at the field value $\varphi_*$ as functions of $\varphi_\text{end}/M_P$. The dotted, dashed and solid lines correspond to the values $N_*=40$, 50 and 60, respectively.\footnote{
We have chosen the highest end point of the horizontal axis of Fig.~\ref{slow roll vs phi_end} to be $\varphi_\text{end}/M_P=e^{W(1/\sqrt{2})}=1.57$ at which $C=0$  and $\V(\Lambda)=0$. In this case, we cannot connect the potential $\V$ to $\V_\text{SM}$, even if the latter were zero at $\Lambda$.
}

\begin{figure}[tn]
\begin{center}
\hfill
\includegraphics[width=.3\textwidth]{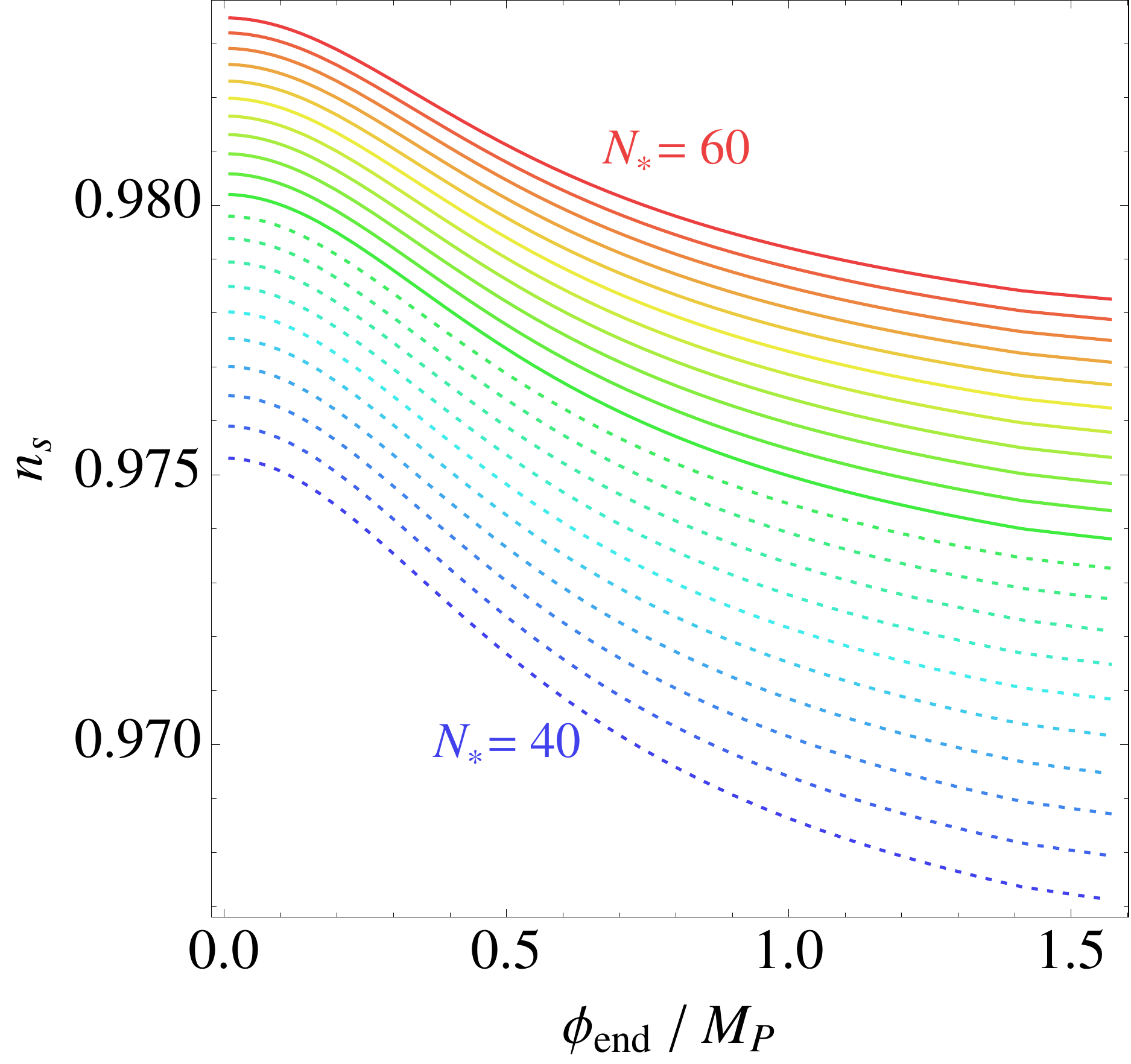}
\hfill
\includegraphics[width=.3\textwidth]{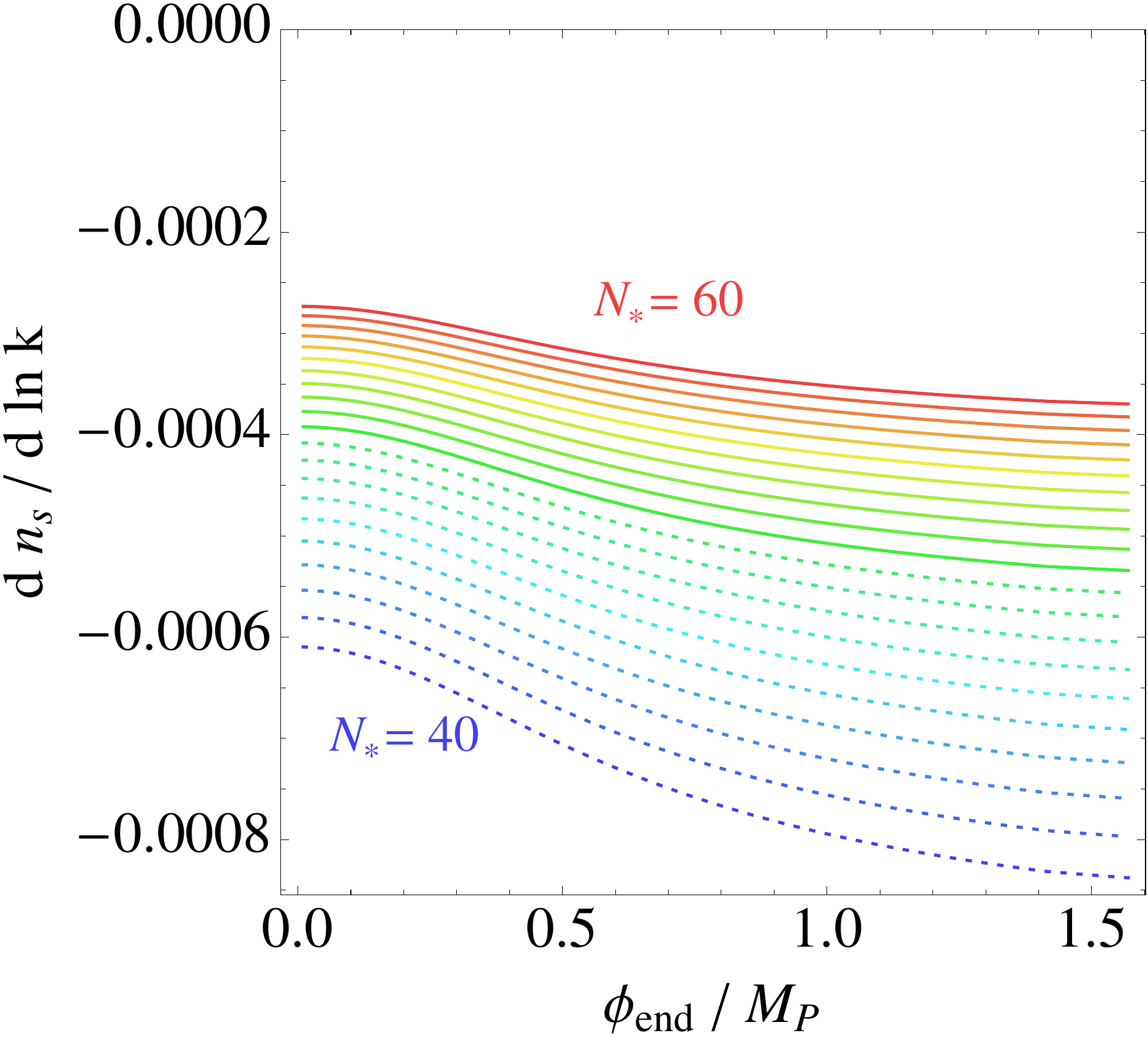}
\hfill
\includegraphics[width=.32\textwidth]{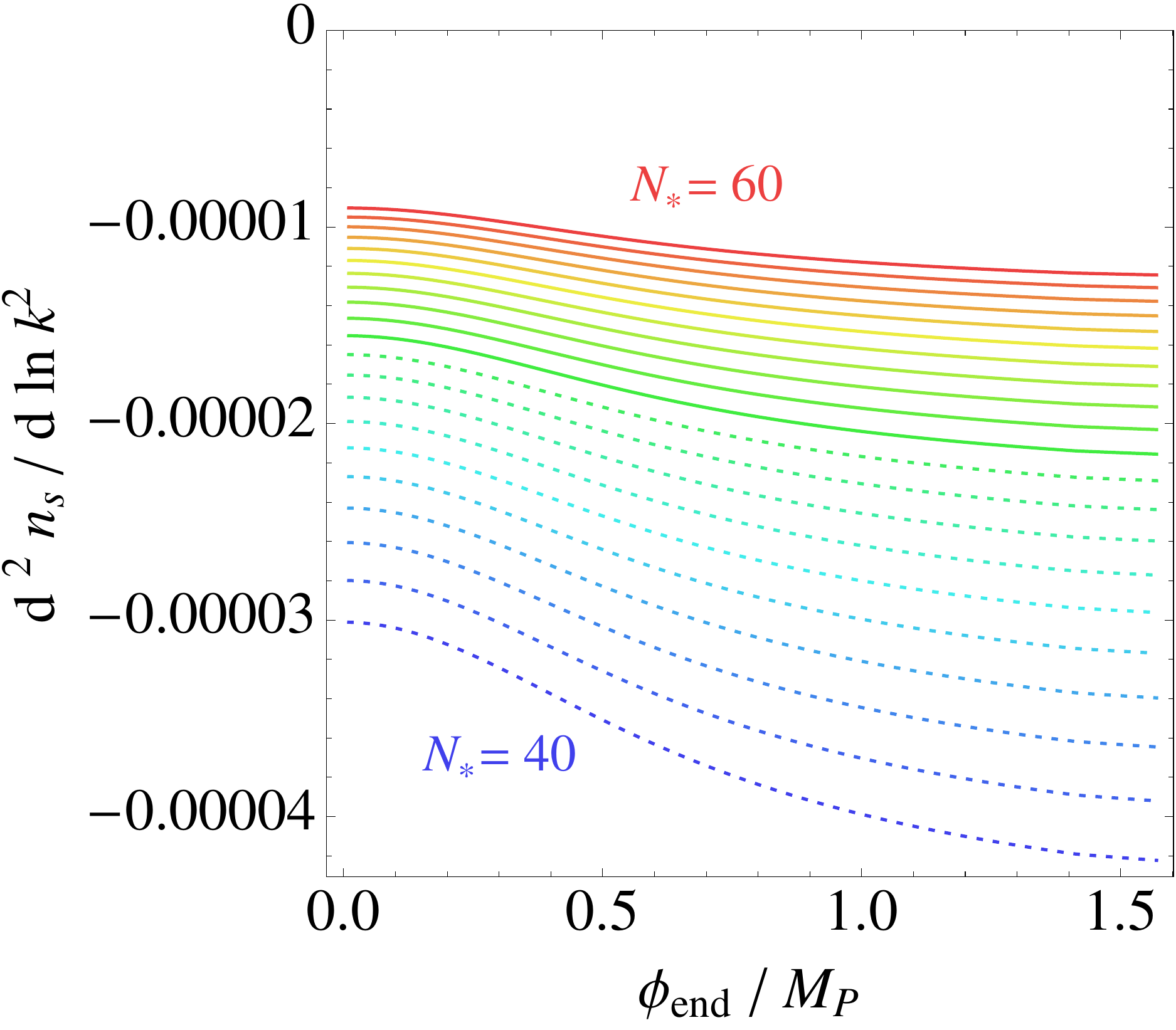}
\hfill\mbox{}\bigskip\\
\hfill
\includegraphics[width=.3\textwidth]{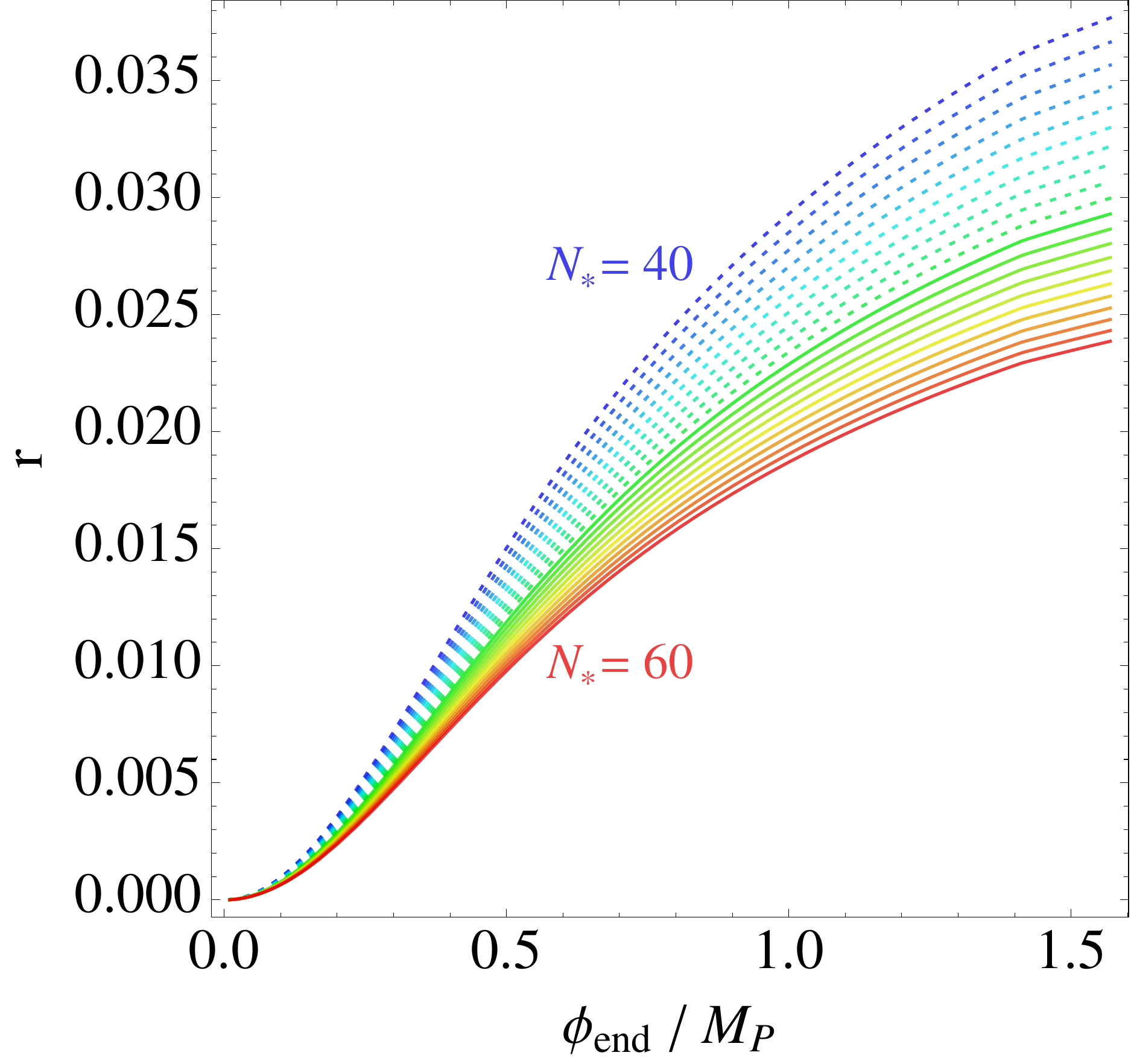}
\hfill
\includegraphics[width=.3\textwidth]{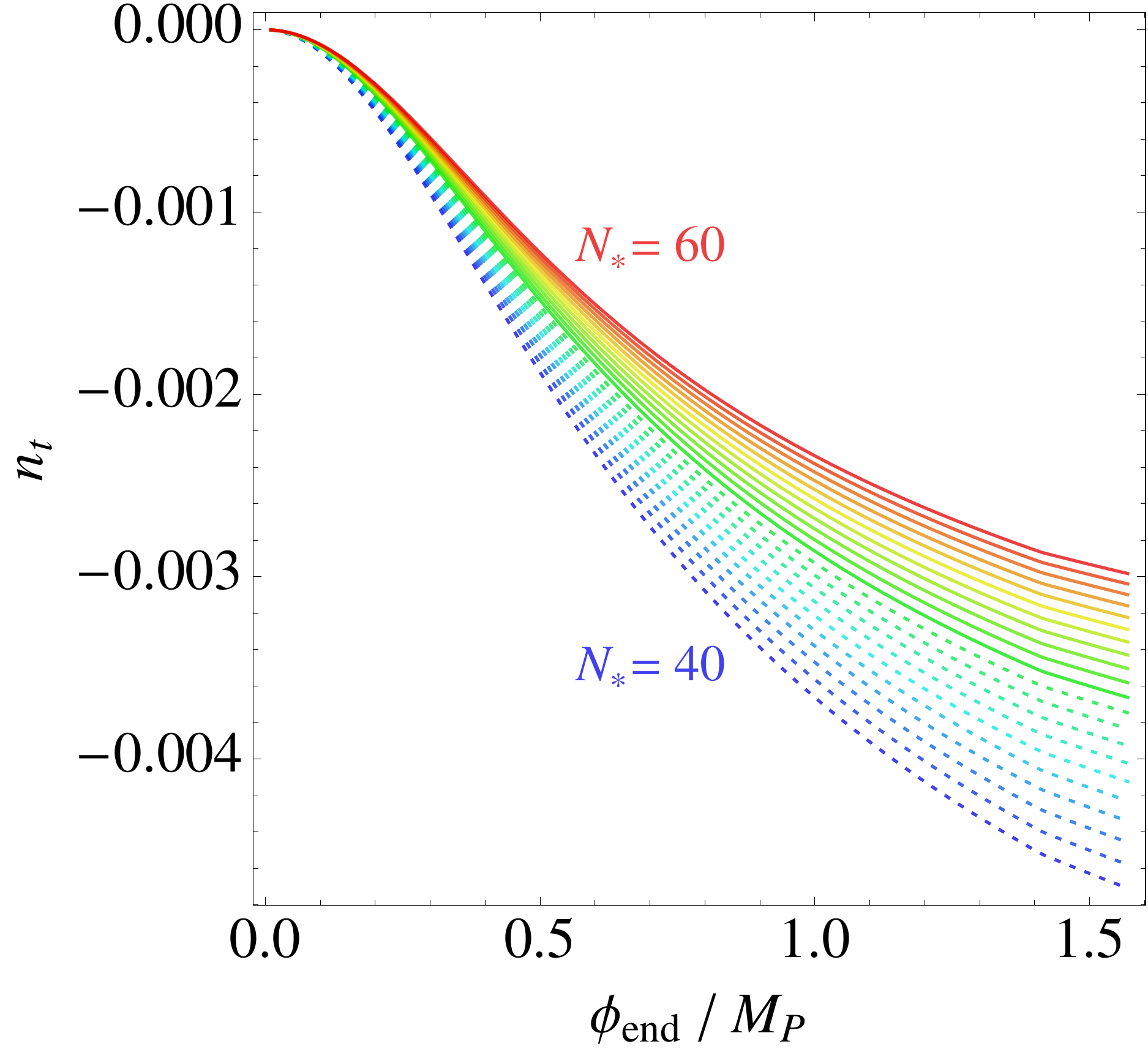}
\hfill
\includegraphics[width=.3\textwidth]{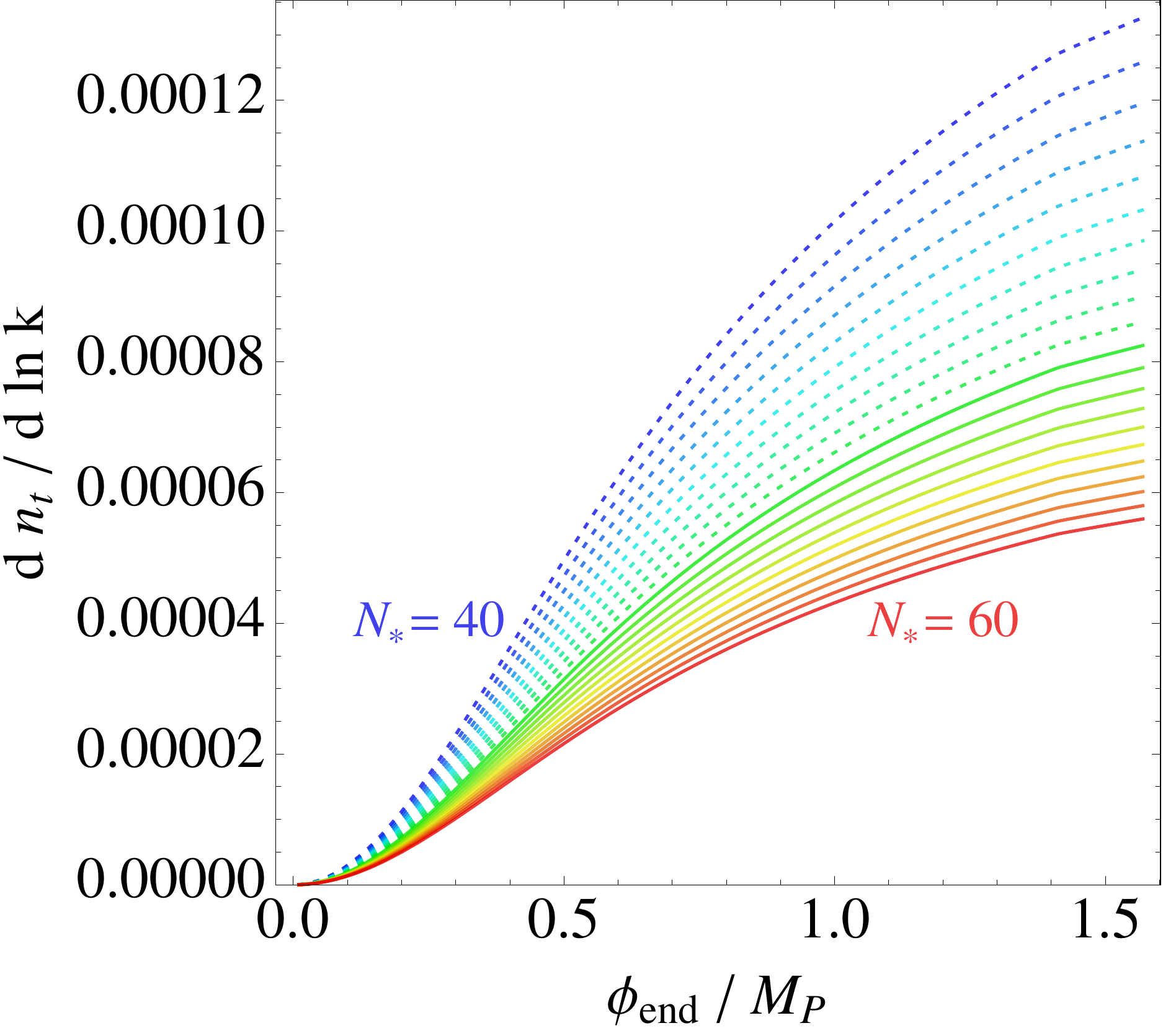}
\hfill\mbox{}
\caption{$n_s$, ${\df n_s\over\df\ln k}$, ${\df^2n_s\over\df\ln k^2}$, $r$, $n_t$, and ${\df n_t\over\df\ln k}$ as functions of $\varphi_\text{end}/M_P$.
}\label{indices vs phi_end}
\end{center}
\end{figure}
Once the slow roll parameters are given, the spectral indices, their running, and their running of running are completely fixed. In Fig.~\ref{indices vs phi_end}, we plot them as functions of $\varphi_\text{end}$. The solid (dotted) lines represent the values for $N_*$ from 50 to 60 (40 to 49). The values for $N_*$ below 50 are just for reference; e.g.\ the late time thermal inflation~\cite{Lyth:1995ka} can reduce the corresponding $N_*$ \rev{to the observed value of~$k_*$}; see Ref.~\cite{Liddle:2003as} for related discussions. When we vary $N_*$ and $\varphi_\text{end}$ within the ranges $50\leq N_*\leq 60$ and $0<\varphi_\text{end}<1.57M_P$, we get
\al{
0.980\text{--}0.984
	&>n_s
			>0.974\text{--}0.979,	&
0	&<	r	<0.029\text{--}0.024,	\nn
\cmag{-}(4.0\text{--}2.7)\times10^{-4}
	&\cmag{>}	{\df n_s\over\df\ln k}
			\cmag{> } \cmag{ -}(5.3\text{--}3.7)\times10^{-4},	&
0
	&>	n_t	>-(3.7\text{--}3.0)\times10^{-3},	\nn
-(1.6\text{--}0.9)\times10^{-5}
	&>	{\df^2n_s\over{\df\ln k}^2}
			>-(2.2\text{--}1.2)\times10^{-5},	&
0	&<	{\df n_t\over\df\ln k}
			<(8.2\text{--}5.6)\times10^{-5}\rev{,}
			\label{numbers shown}
}
\rev{where the order of the inequality corresponds to that of $0<\varphi_\text{end}<1.57M_P$; the range of numbers denoted by the en-dash ``--'' corresponds to the range $N_*=50\text{--} 60$.}

\begin{figure}[tn]
\begin{center}
\hfill
\includegraphics[width=.3\textwidth]{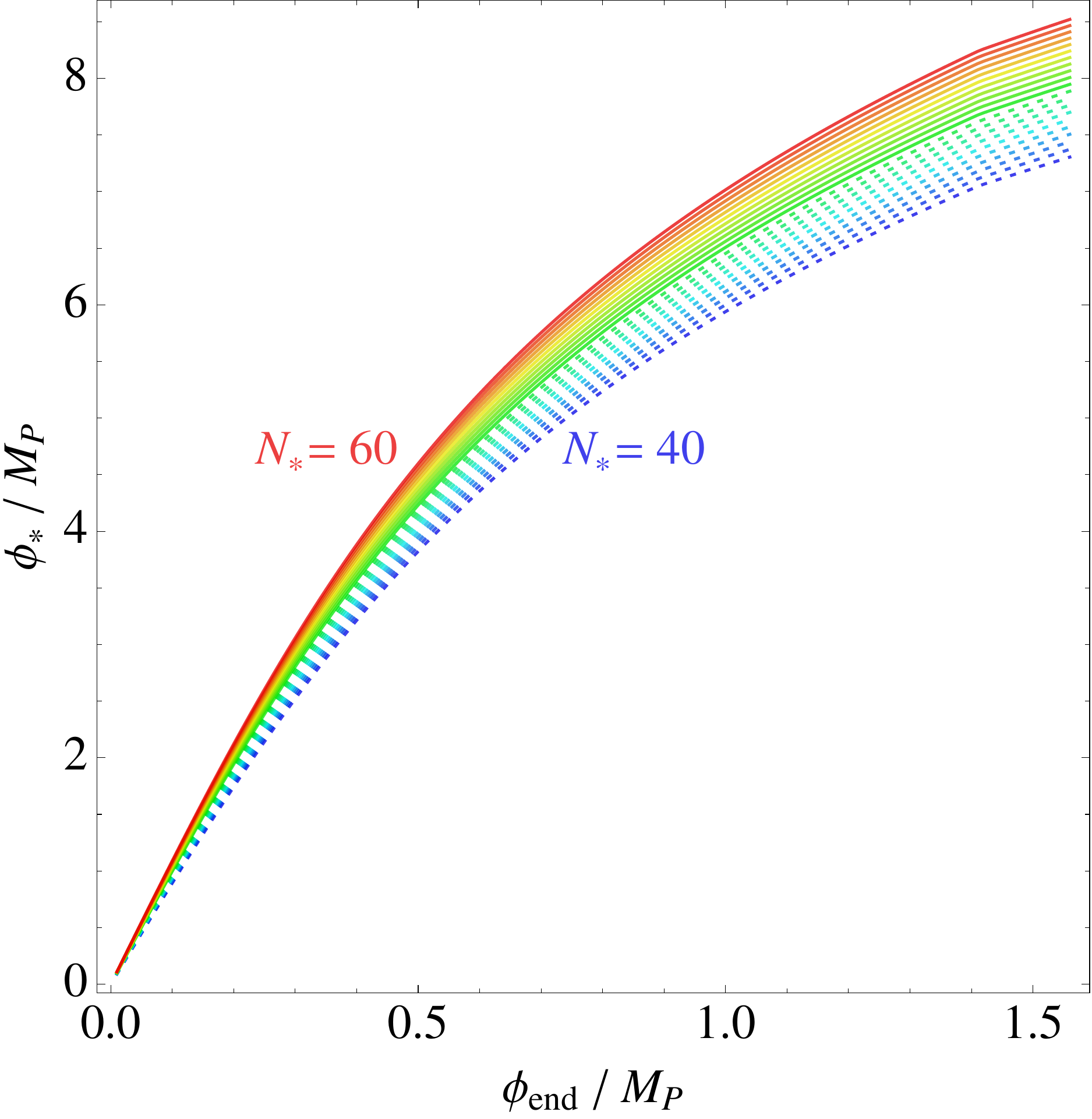}
\hfill
\includegraphics[width=.35\textwidth]{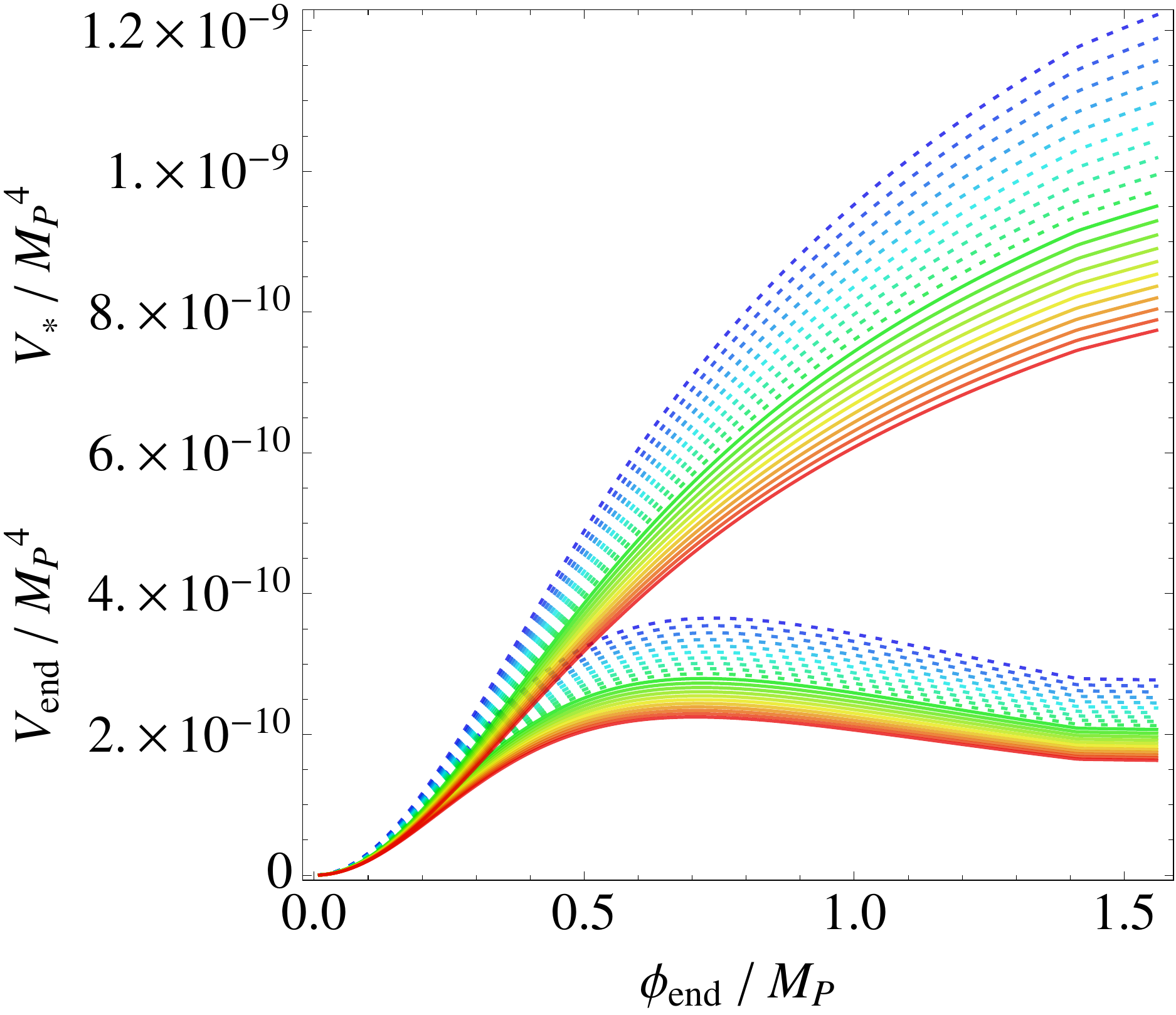}
\hfill\mbox{}
\caption{Left: $\varphi_*/M_P$ as a function of $\varphi_\text{end}/M_P$. Right: Potential values, $\V_*=\V(\varphi_*)$ and $\V_\text{end}=\V(\varphi_\text{end})$, as functions of $\varphi_\text{end}/M_P$. The larger (smaller) values correspond to $\V_*/M_P^4$ ($\V_\text{end}/M_P^4$). We indicate $N_*$ the same as in Fig.~\ref{indices vs phi_end}.
}\label{functions of phi_end}
\end{center}
\end{figure}

So far we have not considered $\V_1$, since it is sufficient to fix $C$ to determine the slow-roll parameters.
Now we determine $\V_1$ by the magnitude of the density perturbation~\eqref{overall normalization}:
\al{
\V_1
	&=	12\pi^2A_sM_P^4
		\paren{M_P\over\varphi_*}^2\paren{1\over C+\ln{\varphi_*\over M_P}}^3.
}
Then the potential and its derivatives at an $e$-folding $N_*$ are completely fixed.
In Fig.~\ref{functions of phi_end}, we plot $\varphi_*$, $\V_\text{end}$, and $\V_*$ as functions of $\varphi_\text{end}$. We indicate $N_*$ the same as in Fig.~\ref{indices vs phi_end}.

\begin{figure}[tn]
\begin{center}
\hfill
\includegraphics[width=.5\textwidth]{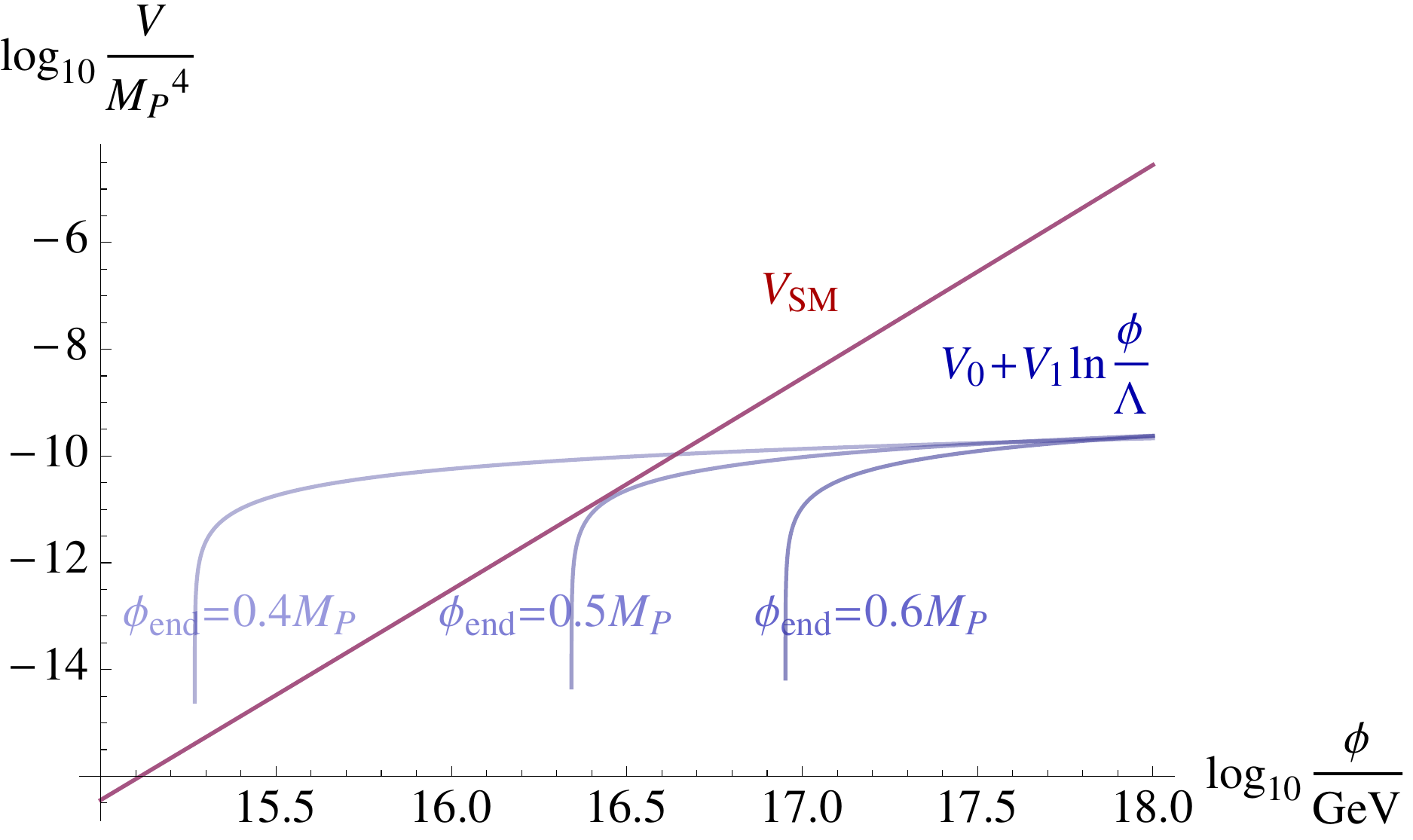}
\hfill\mbox{}
\caption{Three blue curves are high energy potential~\eqref{log potential} for $\varphi_\text{end}=0.4M_P$, $0.5M_P$, and $0.6M_P$ with $N_*=50$. Red line is the SM potential~\eqref{Higgs potential} with the top quark mass $M_t=170.5\GeV$.}\label{V vs phi}
\end{center}
\end{figure}
For a given $\varphi_\text{end}$ we have obtained the constants $\V_1$ and $C$. If we demand that the high scale potential~\eqref{log potential}, fixed by these values, is directly connected with the SM potential at $\Lambda$, then we can fix $\Lambda$ by
\al{
\V_1\paren{C+\ln{\Lambda\over M_P}}
	&=	\V_\text{SM}(\Lambda).
		\label{matching}
}
In Fig.~\ref{V vs phi} we present left and right hand sides of Eq.~\eqref{matching} for $N_*=50$ and $M_t=170.5\GeV$ to illustrate the situation. We see that the low energy SM potential can be directly connected to the high energy one when and only when $\varphi_\text{end}\lesssim 0.5M_P$. This critical value of $\varphi_\text{end}$ is not sensitive to the choice of $N_*$ and $M_t$.

\begin{figure}[tn]
\begin{center}
\hfill
\includegraphics[width=.3\textwidth]{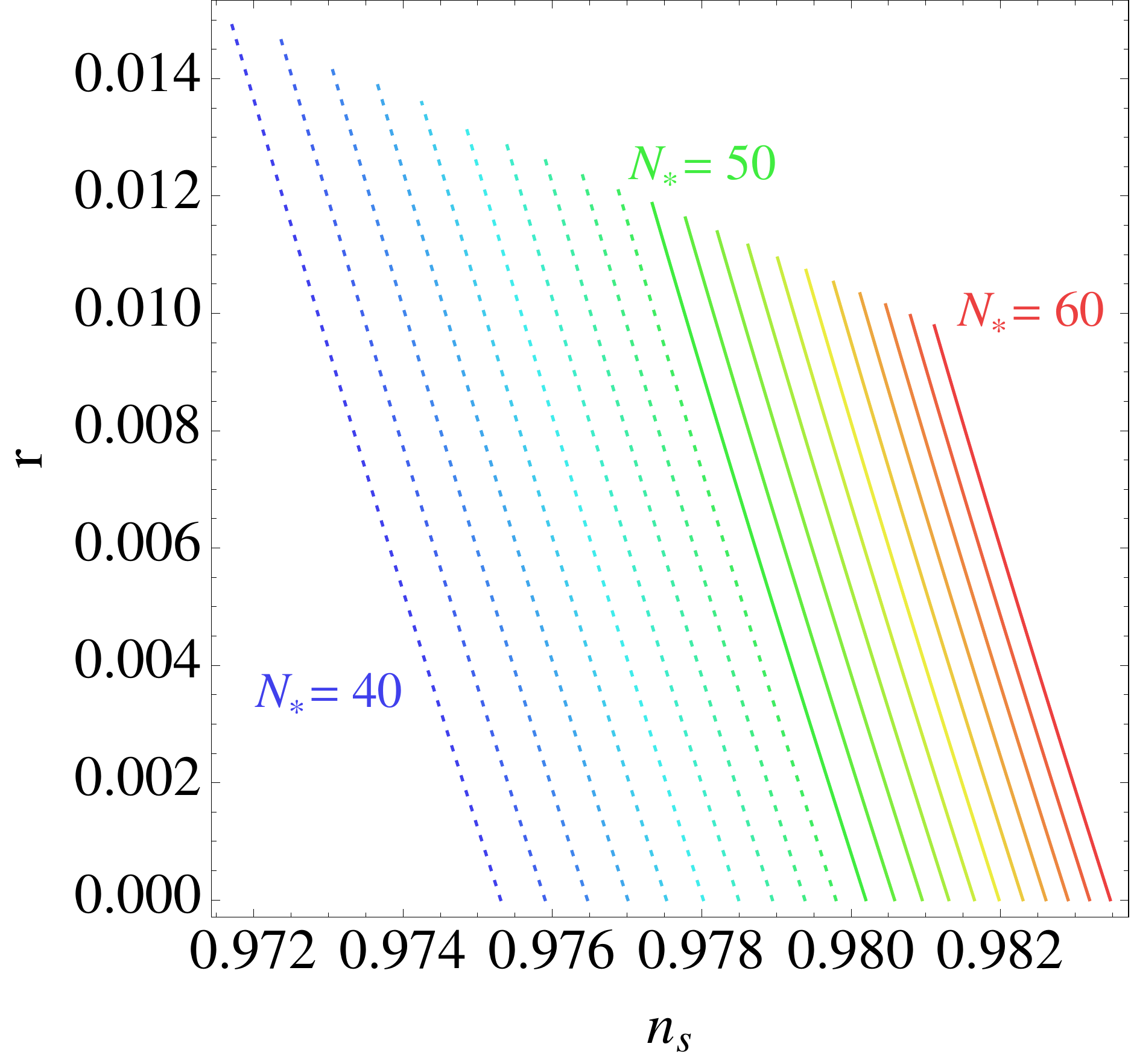}
\hfill
\includegraphics[width=.3\textwidth]{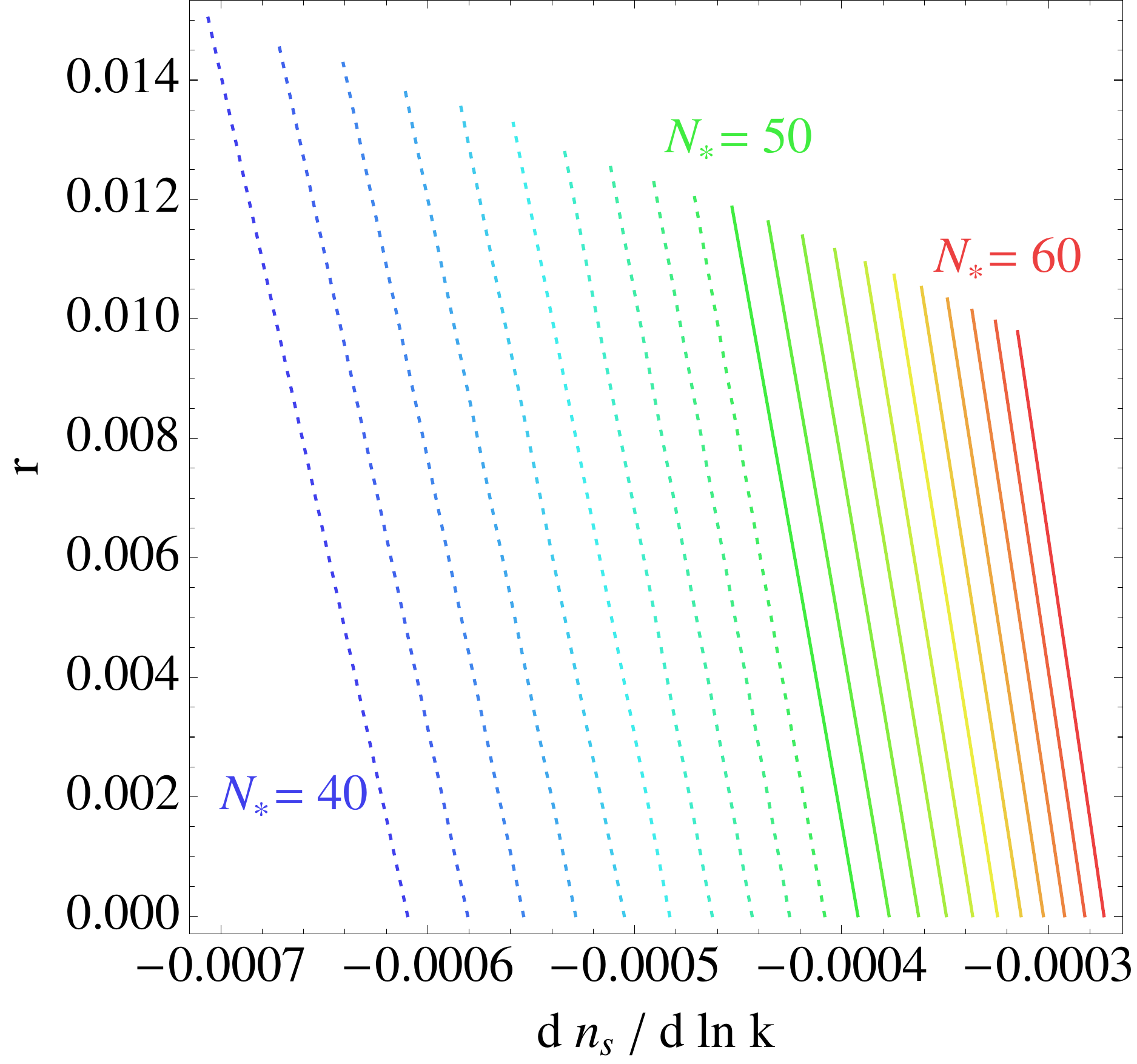}
\hfill\mbox{}\bigskip\\
\hfill
\includegraphics[width=.3\textwidth]{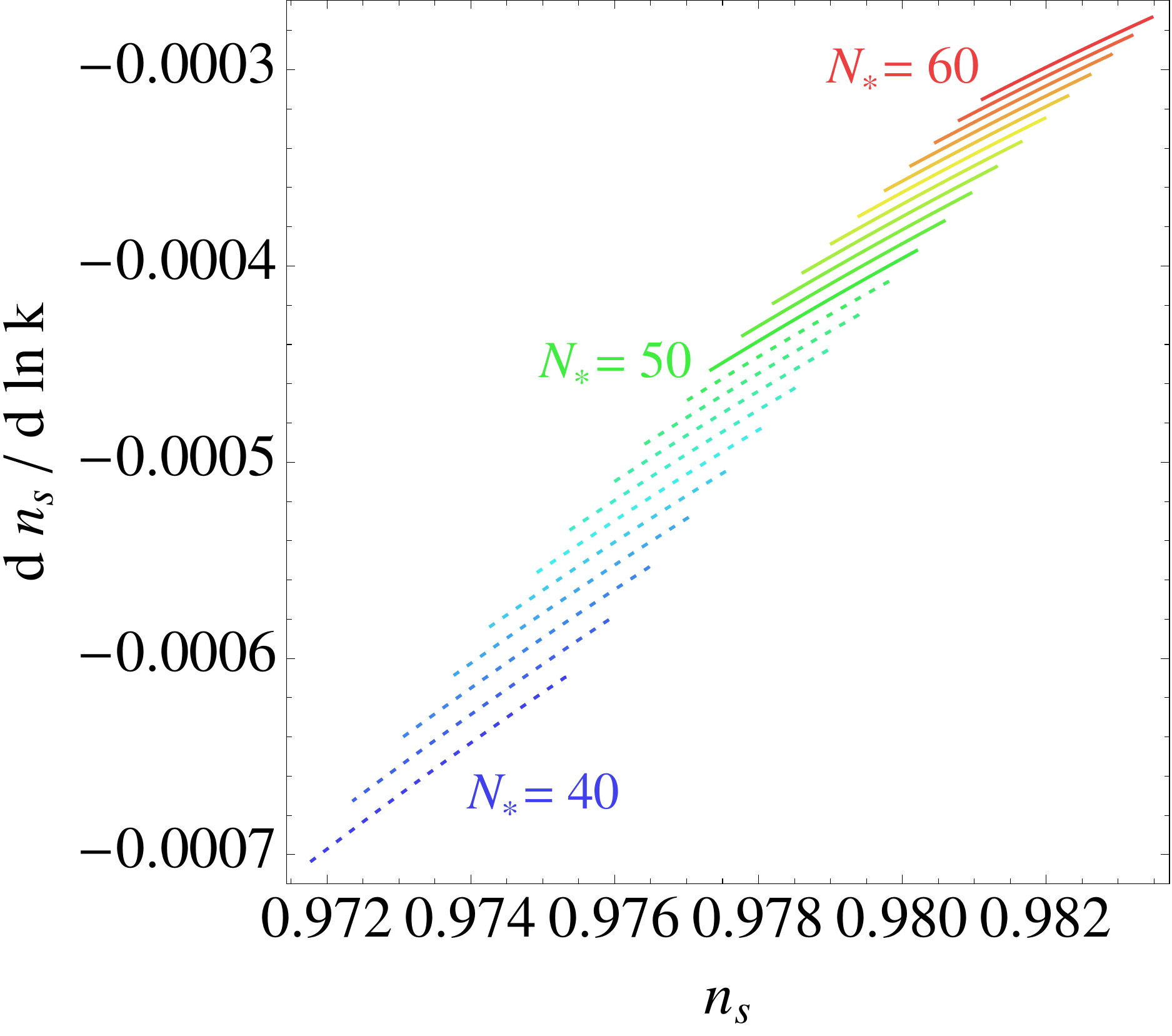}
\hfill
\includegraphics[width=.34\textwidth]{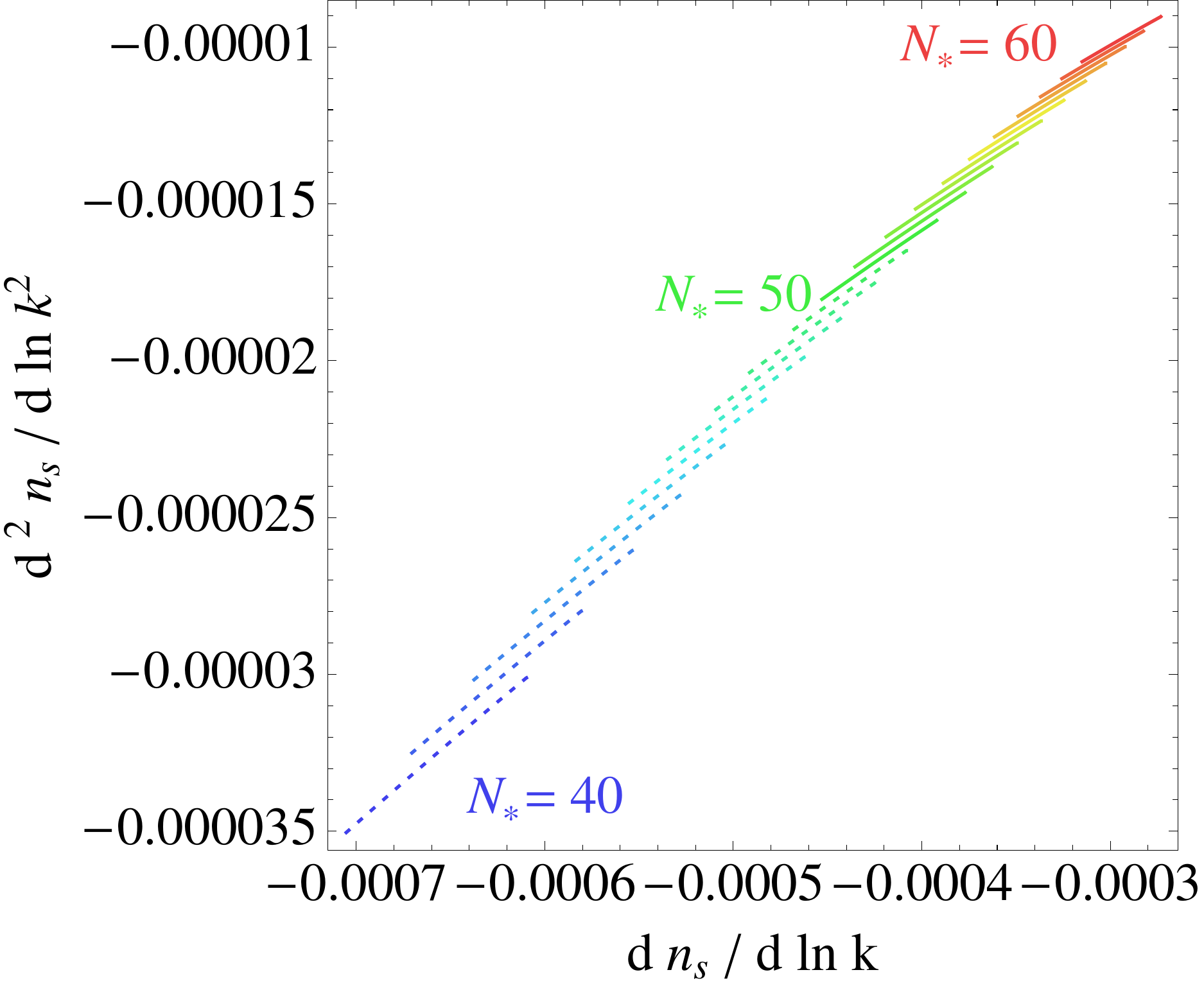}
\hfill\mbox{}
\caption{
At upper-left, upper-right, lower-left, and lower-right, respectively, we have plotted
$r$ vs $n_s$, $r$ vs ${\df n_s\over\df\ln k}$, ${\df n_s\over\df\ln k}$ vs $n_s$, and ${\df^2 n_s\over\df\ln k^2}$ vs ${\df n_s\over\df\ln k}$ for the potential~\eqref{log potential} with $0<\varphi_\text{end}<0.5M_P$.}\label{r vs ns}
\end{center}
\end{figure}
Then we plot $r$ vs $n_s$, $r$ vs ${\df n_s\over\df\ln k}$, ${\df n_s\over\df\ln k}$ vs $n_s$, and ${\df^2 n_s\over\df\ln k^2}$ vs ${\df n_s\over\df\ln k}$ for $0<\varphi_\text{end}<0.5M_P$ in Fig.~\ref{r vs ns}, which can be compared with Figs.~1--5 in Ref.~\cite{Ade:2013uln}.
When we vary $N_*$ and $\varphi_\text{end}$ within the ranges $N_*=50\text{--} 60$ and $0<\varphi_\text{end}<0.5M_P$, we get
\al{\label{predictions}
0.980\text{--}0.983
	&>n_s
			>0.977\text{--}0.981,	&
0	&<	r	<0.012\text{--}0.010,	\nn
\cmag{-}(3.9\text{--}2.7)\times10^{-4}
	&\cmag{>}	{\df n_s\over\df\ln k}
			\cmag{>} \cmag{-}(4.5\text{--}3.1)\times10^{-4},	&
0
	&>	n_t	>-(1.5\text{--}1.2)\times10^{-3},	\nn
-(1.6\text{--}0.9)\times10^{-5}
	&>	{\df^2n_s\over{\df\ln k}^2}
			>-(1.8\text{--}1.0)\times10^{-5},	&
0	&<	{\df n_t\over\df\ln k}
			<(3.1\text{--}2.2)\times10^{-5} \rev{,}
}
\rev{where the order of the inequality corresponds to that of $0<\varphi_\text{end}<0.5M_P$; see Eq.~\eqref{numbers shown}.}
From Fig.~\ref{indices vs phi_end}, we see that we need rather small $N_*\sim40$ \rev{for $0<\varphi_\text{end}<0.5M_P$ in order} to account for the observed value of~$n_s$ \rev{in Eq.~}\eqref{tensor to scalar ratio}.

\begin{figure}[tn]
\begin{center}
\hfill
\includegraphics[width=.4\textwidth]{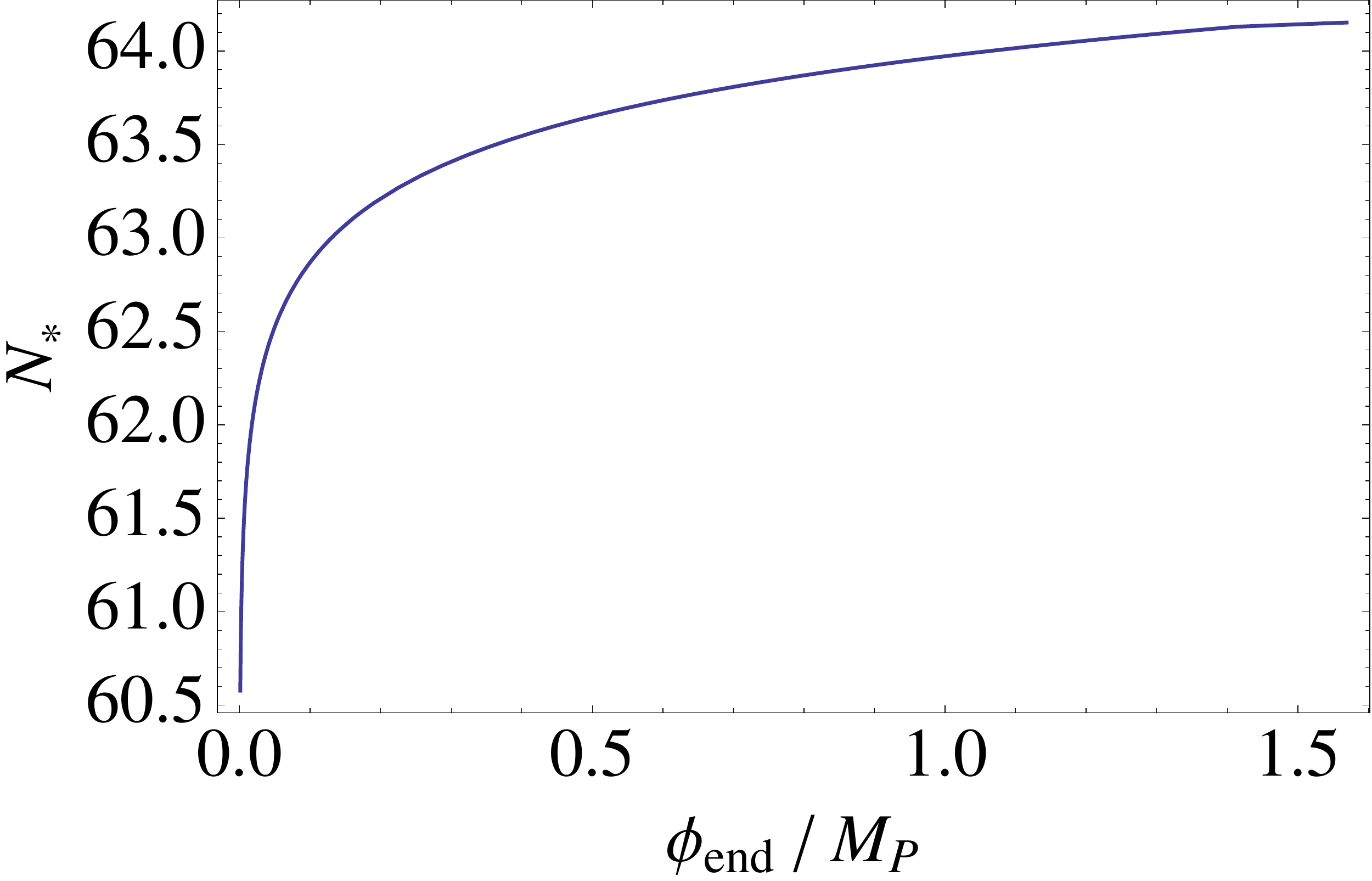}
\hfill\mbox{}
\caption{
\rev{The $e$-foldings $N_*$ as a function of $\varphi_\text{end}/M_P$ corresponding to the pivot scale $k_*=0.002\,\text{Mpc}^{-1}$ in the instantaneous decay approximation; see text.}
}\label{Nstar}
\end{center}
\end{figure}
\rev{
We note that a large field inflation with scale $\gtrsim 10^{17}\GeV$ tends to require relatively high value of $N_*$, barring the late-time thermal inflation mentioned above.
When we approximate that the Higgs field decays into the SM modes instantaneously after the inflation~\cite{Bezrukov:2007ep}, the reheating temperature is given by\footnote{
\rev{
We note that the scale $\varphi_\text{end}$ is beyond $\Lambda$, which is the UV cutoff scale of the SM.
Generically one expects that, above $\Lambda$, there appears extra degrees of freedom besides the SM modes.
In general, these modes, say excited string modes in string theory, decay with the rate of the order of $\Lambda$ times a coupling constant, and we assume that these modes in the decay chain does not affect the result very much. 
}
}
\al{
{\pi^2\over30}g_*T_\text{reh}^4
	&=	\V_\text{end},
}
where $g_*\simeq106.75$ is the effective number of degrees of freedom in the SM;
the resultant reheating temperature is $T_\text{reh}\simeq4\times10^{15}\GeV$ for $\varphi_\text{end}=0.5M_P$.
Then the $e$-folding number, corresponding to the pivot scale $k_*=0.002\,\text{Mpc}^{-1}$ and the Hubble parameter $H_0=67.3\,\text{km}\,\text{s}^{-1}\text{Mpc}^{-1}$, is given by~\cite{Ade:2013uln}:
\al{
N_*	&\simeq
	69+{1\over4}\ln{\V_*\over M_P^4}+{1\over4}\ln{\V_*\over\V_\text{end}}.
}
\addedSecond{Using the value of $\V_*$ and $\V_{\text{end}}$ as depicted in Fig.~\ref{functions of phi_end}}
, we get the $e$-folding number as a function of $\varphi_\text{end}$, which is plotted in Fig.~\ref{Nstar}. \addedSecond{This indicates that the large field inflation requires $N_*\gtrsim60$.}
}

\rev{
\addedSecond{On the other hand, as we have seen around Eq.~\eqref{matching} and Eq.~\eqref{predictions}, we need a small value of $N_*\sim40$, if we want to directly connect the log potential with the SM one. To do so, $\varphi_{\text{end}}$ needs relatively small, $\varphi_{\text{end}}<0.5 M_P$, and $N_*$ should be around $40$ in order to obtain a realistic value of $n_s$.}



There are two ways to solve this apparent inconsistency. \addedSecond{One} is to note that}
indeed \addedSecond{we do not have} to connect the log potential to $\V_\text{SM}$ so strictly, as \addedSecond{it is unclear} what happens around $\varphi\sim\Lambda$.\footnote{
\rev{
For example, the Coleman-Weinberg potential that has an explicit momentum cutoff $\Lambda$ turns into the log-type potential only at large field values $\varphi\gg\Lambda$ as shown in Appendix~\ref{log motivation}.
}
}
\rev{All we need is that the end point value of the inflaton potential is larger than the SM potential at its UV cutoff scale: $\V_\text{end}>\V_\text{SM}(\Lambda)$ for $\varphi_\text{end}>\Lambda$.}
Then we can \rev{take} larger $\varphi_\text{end}$ to obtain smaller value of $n_s$. 

\begin{figure}[tn]
\begin{center}
\hfill
\includegraphics[width=.5\textwidth]{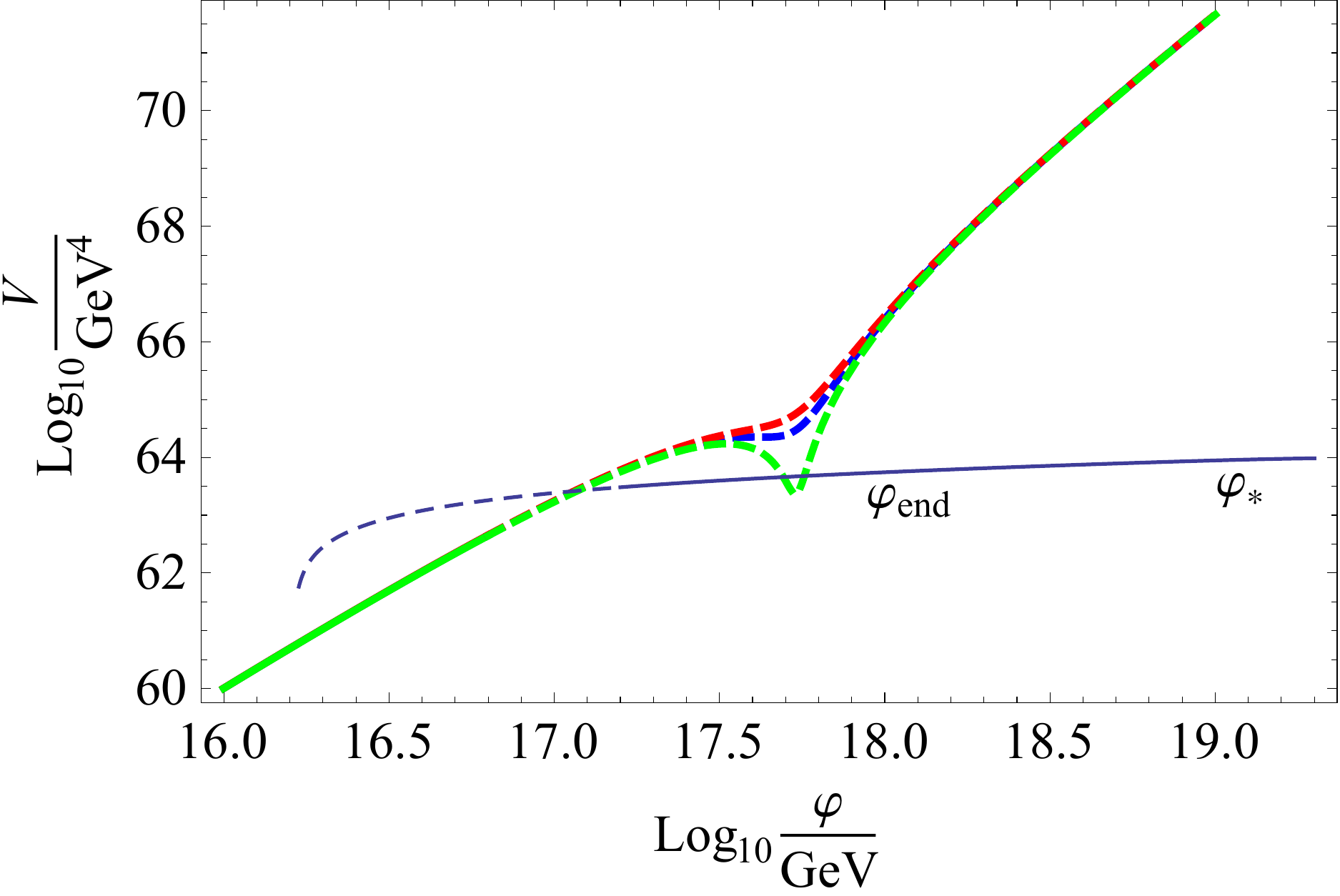}
\hfill\mbox{}
\caption{
\rev{
Log potential~\eqref{log potential R} plus the correction~\eqref{potential correction} as a function of $\varphi$ is drawn in the log-log plot as the solid line in the right, which \addedSecond{is the modification of} the line for $\varphi_\text{end}=0.5M_P$ in Fig.~\ref{V vs phi}; $\varphi_*$ and $\varphi_\text{end}$ are also indicated. The SM lines are the same as in Fig.~\ref{flat potential} but the SM cutoff is taken to be slightly smaller $\Lambda\simeq 10^{17}\GeV$.
}
}\label{matching figure}
\end{center}
\end{figure}
\rev{
 \addedSecond{The other is adding a small correction to the log potential:
\al{
\Delta\V
	&=	\V_1\paren{c_1{\varphi\over M_P}+c_2{\varphi^2\over M_P^2}+\cdots}.
		\label{potential correction}
}
For example, if we choose $\V_1=4\times10^{-11}M_P^4$, $C=5$, $c_1=0.1$, $c_2=-0.01$, and $c_n=0$ for $n\geq3$, then we get $\varphi_\text{end}=0.48\,M_P$, $\varphi_*=4.7M_P$, $N_*=64$, $r=0.008$ and $n_s=0.978$. The resultant potential is illustrated in Fig.~\ref{matching figure}.
}

}

\section{Summary \cred{and discussions}}
The Higgs potential in the Standard Model (SM) can have a saddle point around $10^{17}\GeV$, and its height is suppressed because the Higgs quartic coupling becomes small.
These facts suggest that the SM Higgs field may serve as an inflaton, without assuming the very large coupling to the Ricci scalar of order $10^4$, which is necessary in the ordinary Higgs inflation scenario. In this paper, we have pursued the possibility that the Higgs potential becomes almost flat above the UV cutoff~$\Lambda$.

Since a first order phase transition at the end of the inflation leads to the graceful exit problem, the Higgs potential must be monotonically increasing in all the range below and above $\Lambda$.
From this condition, we get an upper bound on $\Lambda$ to be of the order of $10^{17}\GeV$. 

We have briefly sketched the possible log type potential above~$\Lambda$. We present the motivation of this shape in Appendix~\ref{log motivation}, that is, the Coleman-Weinberg one-loop effective potential becomes of this type above $\Lambda$ when the momentum integral is cut off by $\Lambda$.
The predictions on the parameters of the cosmic microwave background at the $e$-foldings $N_*=50$--60 are: 
the scalar spectral index, $0.980\text{--}0.983>n_s>0.977\text{--}0.981$;
the tensor to scalar ratio, $0<r<0.012\text{--}0.010$;
the running scalar index, 
\cmag{$-\paren{4.5\text{--}3.1}\times10^{-4}<\df n_s/\df\ln k<-(3.9\text{--}2.7)\times10^{-4}$};
the running of running scalar index, $-\paren{1.6\text{--}0.9}\times10^{-5}>\df^2n_s/{\df\ln k}^2>-(1.8\text{--}1.0)\times10^{-5}$;
the tensor spectral index, $0>n_t>-(1.5\text{--}1.2)\times10^{-3}$; and
the running tensor index, $0<\df n_t/\df\ln k<(3.1\text{--}2.2)\times10^{-5}$.

\cred{
In this paper, we have pursued the bottom-up approach from the latest Higgs data, without assuming any other structure than the SM below~$\Lambda$. 
We have shown in Fig.~\ref{topconstraint} that even if we allow arbitrary potential above $\Lambda$, still the restriction is rather severe to achieve this minimal Higgs inflation scenario. 
}

It is curious that the upper bound on $\Lambda$ from the minimal Higgs inflation coincides with the scale where \cred{the quartic coupling} $\lambda$ and its beta function (and possibly the bare Higgs mass) vanish. This coincident scale $\sim10^{17}\GeV$ is close to the string scale in the conventional perturbative superstring scenario.\footnote{
\cred{See e.g.\ Refs.~\cite{Hebecker:2012qp,Hebecker:2013lha} for trials to explain the smallness of the quartic coupling in string theory context.}
}
This fact may suggest that the physics of the SM, string theory, and the universe are all directly connected. 

\cred{
There are possibilities that realizes the flatness from the gauge symmetry as in the gauge-Higgs unification scenario~\cite{Haba:2005kc,Haba:2008dr,Maru:2013ooa,Maru:2013bja,Maru:2013qla}. See also Refs.~\cite{Hebecker:2012qp,Hebecker:2013lha} for other stringy attempts. Furthermore, Ref.~\cite{Hebecker:2011hk,Hebecker:2012aw} derives the log potential of the type~\eqref{log potential}. It would be interesting to construct a realistic string model that breaks the supersymmetry at string scale\footnote{
\cred{Generally number of superstring vacua that breaks supersymmetry at string scale is much larger than the one that preserves supersymmetry~\cite{Kawai:1985xq,Lerche:1986cx,Antoniadis:1986rn}.}
}
and realizes the flat Higgs potential above $\Lambda$ consistent to the cosmological observations.
One can even go beyond the symmetry argument of the ordinary quantum field theory/string theory to realize the flat potential, such as with the MPP~\cite{Froggatt:1995rt,Froggatt:2001pa,Nielsen:2012pu}, the classical conformality around $\Lambda$~\cite{Meissner:2006zh,Iso:2009ss,Iso:2009nw,Holthausen:2009uc,Iso:2012jn}, the multiverse~\cite{Kawai:2011qb}, the anthropic principle~\cite{Hall:2009nd}, etc.
}

\subsection*{Acknowledgement}
We thank Tetsutaro Higaki for a useful comment. \Magenta{We also thank Finelli Fabio for his kind explanation about Ref.~\cite{Ade:2013uln}.}
This work is in part supported by the Grant-in-Aid for Scientific Research Nos.~22540277 (HK), 23104009, 20244028, and 23740192 (KO) and for the Global COE program ``The Next Generation of Physics, Spun from Universality and Emergence.'' The work of Y. H. was supported by a Grant-in-Aid for  Japan Society for the Promotion of Science (JSPS) Fellows No.25$\cdot$1107. 

\appendix
\section*{Appendix}

\begin{figure}[tn]
\begin{center}
\hfill
\includegraphics[width=.4\textwidth]{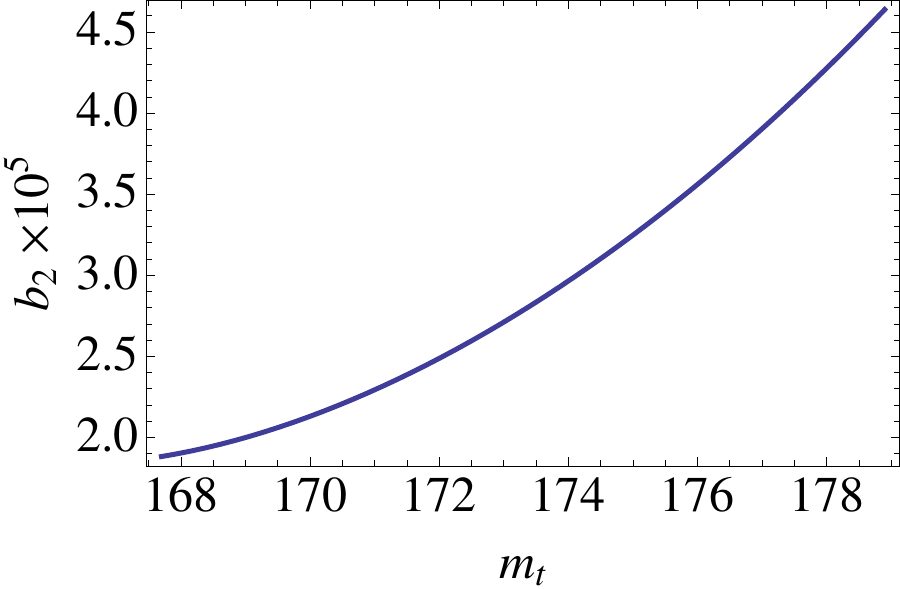}
\hfill\mbox{}
\caption{Running of the beta function $b_2$ at the scale $\Lambda_0$ of vanishing beta function as a function of top pole mass $M_t$. 
}\label{b2}
\end{center}
\end{figure}

\section{SM Higgs as inflaton}\label{SM Higgs as inflaton}
To see more explicitly the argument shown in Sec.~\ref{Higgs potential in SM} for the difficulty in directly using the SM Higgs field as an inflaton, we first show how to make a saddle point.
Let us expand the potential around the point $\varphi_0$ ($\sim 10^{17}\GeV$) of vanishing beta function $\beta_\lambda=0$:
\al{
\left.\V\right|_{\varphi\sim\varphi_0}
	&=	{\varphi^4\over4}\sqbr{
			\lambda_0
			+b_2\paren{\ln{\varphi\over\varphi_0}}^2
			+b_3\paren{\ln{\varphi\over\varphi_0}}^3
			+\cdots
			},
			\label{approximated potential}
}
where $b_i$ are given by
\al{
b_2	&=	{1\over2}{\df^2\lambda\over{\df\ln\mu}^2}
	=	{1\over2}\sum_i\beta_i{\partial\beta_\lambda\over\partial\lambda_i},	&
b_n	&=	{1\over n!}{\df^n\lambda\over{\df\ln\mu}^n}
	=	O\fn{\paren{16\pi^2}^{-n}},
}
with $\lambda_i$ representing~\cite{Hamada:2012bp} the Yukawa coupling squared, $y_t^2$ etc., the gauge coupling squared, $g_Y^2$, $g_2^2$, $g_3^2$, and the quartic coupling $\lambda$.
Note that each $\beta_i=\df\lambda_i/\df\ln\mu$ has a loop suppression factor $1/16\pi^2$.
We see that the SM Higgs potential can always have a saddle point by choosing a particular value of $\lambda_0$ by adjusting the top quark mass. For example, when we approximate $b_3=0$, the saddle point is realized at $\varphi=e^{-1/4}\varphi_0$ by choosing $\lambda_0=b_2/16$. In the SM, $b_2$ takes values $b_2\simeq(1.9\text{--}4.6)\times10^{-5}$ for the 95\% confidence interval from the top quark mass~\eqref{Djouadi mass}, see Fig.~\ref{b2}.

In order to show the difficulty of the inflection point inflation with the SM Higgs potential, it suffices to expand the potential around its inflection point $\varphi_c$ that satisfies $\V''_c:=\V_{\varphi\varphi}(\varphi_c)=0$:
\al{
\V(\varphi)
	&=	\V_c
		+\V_c'\,\paren{\varphi-\varphi_c}
		+{\V_c'''\over3!}\paren{\varphi-\varphi_c}^3
		+\cdots,
}
where $\V_c''':=\V_{\varphi\varphi\varphi}(\varphi_c)$ and we tune the top quark mass in order to make $\V_c':=\V_\varphi(\varphi_c)$ very small.
The $e$-folding from $\varphi_c+\delta\varphi_*$ to $\varphi_c-\delta\varphi_\text{end}$ becomes
\al{
N_*
	&=	\sqrt{2\over \V_c'\V_c'''}{\V_c\over M_P^2}\sqbr{
			\arctan\fn{\sqrt{\V_c'''\over2\V_c'}\delta\varphi_\text{end}}
			+\arctan\fn{\sqrt{\V_c'''\over2\V_c'}\delta\varphi_*}
			}\nn
	&\quad
		+{2\V_c'\over3M_P^2\V_c'''}\ln{2\V_c'+\V_c'''\delta\varphi_*^2\over2\V_c'+\V_c'''\delta\varphi_\text{end}^2}
		+{\delta\varphi_*^2-\delta\varphi_\text{end}^2\over6M_P^2}.
		\label{e-folding for II}
}
In the following, we discuss in detail the three cases that are sketched in the text:
\begin{itemize}
\item First possibility is to put $\V_c'=0$ and earn the $e$-folding near the saddle point. The $e$-folding~\eqref{e-folding for II} for $\delta\varphi_\text{end}>0$ and $\delta\varphi_*<0$ becomes
\al{
N_*	&=	{2\V_c\over M_P^2\V_c'''}\paren{
			{1\over\ab{\delta\varphi_*}}
			-{1\over\delta\varphi_\text{end}}
			}
		-{\delta\varphi_\text{end}^2-\delta\varphi_*^2\over6M_P^2}
	\simeq
		{2\V_c\over M_P^2\V_c'''\ab{\delta\varphi_*}}.
			\label{e-folding}
}
Close to the saddle point, we have
\al{
\epsilon_V
	&=	{M_P^2\paren{\V_c'''}^2\over8\V_c^2}\delta\varphi_*^4
		+O\fn{\delta\varphi_*^7}.
		\label{epsilonV}
}
Putting Eq.~\eqref{e-folding} into Eq.~\eqref{epsilonV}, the slow roll parameter reads
\al{
\epsilon_V={2\V_c^2\over N_*^4M_P^6\paren{\V_c'''}^2},
	\label{epsilonV saddle}
}
and hence the scalar perturbation becomes
\al{
A_s	&=	{N_*^4M_P^2\paren{\V_c'''}^2\over48\pi^2\V_c}.
}
From Eq.~\eqref{approximated potential}, we can compute the values in the SM:
\al{
\V_c'''\sim 10^{-5}\varphi_c\sim 10^{-6}M_P
	\label{third derivative}
}
which results in $A_s\gg1$, far larger than the allowed value~\eqref{As bound}.
\item As a second possibility, one might introduce small $\V_c'\sim10^{-11}M_P^3$ at $\varphi_c$, in order to realize the value of the correct density perturbation at $\varphi_c$. For the SM values $\V_c\sim10^{-9}M_P^4$ and $\V_c'''=10^{-6}M_P$, we obtain from Eq.~\eqref{e-folding for II} the $e$-folding
\al{
N_*\sim 1.
}
This does not work.
\item Finally one may take very tiny $\V_c'$ at $\varphi_c$ to earn enough $e$-folding, in order to realize the inflection point inflation scenario~\cite{Itzhaki:2007nk,Allahverdi:2008bt,Badziak:2008gv,Enqvist:2010vd,Hotchkiss:2011am,Hotchkiss:2011gz,Chatterjee:2011qr,Choudhury:2013iaa,Wang:2013hva,Choudhury:2013jya}, while obtaining necessary amount of $\epsilon_V$ at a point above the inflection point: $\varphi_c+\delta\varphi_*$ with $\delta\varphi_*>0$. In passing through the inflection point $\varphi_c$ from $\varphi_*$ ($>\varphi_c$) to $\varphi_\text{end}$ ($<\varphi_c$), we earn the $e$-folding
\al{
N_* &=	\sqrt{2\pi^2\V_c^2\over M_P^4\V_c'\V_c'''}+O\fn{\paren{\V_c'}^0},
}
and hence we can have as large an $e$-folding as we want by tuning $\V_c'$ small. More concretely, we need
\al{
\V_c'\sim 10^{-15}M_P^3
	\label{first derivative}
}
to get $N_*\sim 50$ with Eq.~\eqref{third derivative}. However, to keep the slow roll parameter
\al{
\eta_V
	&=	{M_P^2\V_c'''\over \V_c}\delta\varphi_*+O\fn{\delta\varphi_*^2}
}
sufficiently small, we need to be close to the inflection point:
\al{
\delta\varphi_*\ll {\V_c\over M_P^2\V_c'''}.
}
Within this range, we get
\al{
\epsilon_V
	&=	{\V_c^2\eta_V^4\over 8M_P^6\paren{\V_c'''}^2}
	\ll	{\V_c^2\over 8M_P^6\paren{\V_c'''}^2}.
		\label{epsilon relation}
}
\begin{itemize}
\item When $\delta\varphi_*^2\gg 2\V_c'/\V_c'''$, we have the same expression as Eq.~\eqref{epsilonV}. From Eq.~\eqref{epsilon relation}, we get
\al{
A_s	&=	{\V_c\over24\pi^2M_P^4\epsilon_V}
	\gg	
		{M_P^2\paren{\V_c'''}^2\over 3\pi^2\V_c}.
}
Putting the SM values, this results in $A_s\gg 10^{-5}$ which is far larger than the observation~\eqref{As bound}.
\item On the contrary when $\delta\varphi_*^2\ll 2\V_c'/\V_c'''$, we get
$\epsilon_V=M_P^2\paren{\V_c'}^2/2\V_c^2$ and hence
\al{
A_s	&	
	=	{\V_c^3\over12\pi^2M_P^6\paren{\V_c'}^2}
	\sim	1,
}
where we have put Eq.~\eqref{first derivative}. We see that this is too large again.
\end{itemize}
\end{itemize}


\section{Motivating log type toy model}\label{log motivation}
We present a motivation for the toy model with the log type potential at $\varphi>\Lambda$:
\al{\label{log potential}
\V	&=	\V_0+ \V_1 \ln \frac{\varphi}{\Lambda}
	=:	\wt\V_0+\V_1\ln{\varphi\over M_P}.
}
In the bare perturbation theory, see e.g.\ Ref.~\cite{Hamada:2012bp}, the one-loop effective potential for the Higgs field $\varphi$ is given by
\al{
\V_\text{eff}(\varphi)
	&=	{m_B^2\over 2}\varphi^2+{\lambda_B\over 4}\varphi^4
		+\sum_i{N_i\over2}\int{\df^4p\over(2\pi)^4}\ln{p^2+c_i\varphi^2\over p^2},
		\label{bare_potential}
}
where the integration is performed over Euclidean four momentum.\footnote{
In Eq.~\eqref{bare_potential}, we have tuned the cosmological constant so that we get $\V_\text{eff}\to0$ as $\varphi\to0$. 
}
Since we are interested in the behavior of the potential at the field value $\varphi$ very much larger than the electroweak scale $\varphi\gg V=246.22\GeV$, we work in the symmetric phase by setting the Higgs VEV to be zero: $V=0$. The number of degrees of freedom, $N_i$, and the coupling to the Higgs, $c_i$, are summarized in Table~\ref{number_table} for species $i$ that have non-negligible coupling to the Higgs. $h$ and $\chi$ are the physical and Nambu-Goldstone modes of the Higgs around the field value $\varphi$, respectively.\footnote{
Though we show our results in the Landau gauge, we can explicitly show that in the $R_\xi$ gauge, depending on the external field $\varphi$, the one-loop result~\eqref{bare_potential} is independent of the gauge parameter $\xi$ if we expand it by $c_i\varphi^2\ll\Lambda^2$.
}
Assuming existence of an underlying gauge invariant regularization, such as string theory, let us cutoff the integral by $\ab{p}<\Lambda$:
\al{
{\df\V_\text{eff}\over\df\varphi^2}
	&=	{1\over2}\sqbr{
			m_B^2
			+\lambda_B\varphi^2
			+\sum_i{N_ic_i\over16\pi^2}\paren{
				\Lambda^2-c_i\varphi^2\ln{\Lambda^2+c_i\varphi^2\over c_i\varphi^2}
				}
			}.
}
The bare Higgs mass $m_B^2$ is tuned to yield the desired value of the low energy mass-squared parameter,
\al{
m_R^2
	&:=	2\left.{\df\V_\text{eff}\over\df\varphi^2}\right|_{\varphi^2\to0},
}
to be zero: $m_R^2=0$, see e.g.\ Ref.~\cite{Hamada:2012bp}.\footnote{
Recall that we are working in the symmetric phase.
}
Then we get
\al{
m_B^2
	&=	-{\Lambda^2\over16\pi^2}\sum_iN_ic_i.
}
To summarize, we generally have
\al{
{\df\V_\text{eff}\over\df\varphi^2}
	&=	{\varphi^2\over2}\sqbr{
			\lambda_B
			-\sum_i{N_ic_i^2\over16\pi^2}
				\ln{\Lambda^2+c_i\varphi^2\over c_i\varphi^2}
			}.
			\label{bare potential final}
}
We see that the bare mass drops out of the effective potential, as it should be.
The form~\eqref{bare potential final} corresponds to the one loop correction to $\lambda_B$.
As a side remark, we comment that the condition $m_B^2=0$ at this one-loop order, namely $\sum_iN_ic_i=0$, is the celebrated Veltman condition~\cite{Veltman:1980mj,Hamada:2012bp}.

Rigorously speaking, the effective potential~\eqref{bare_potential} or \eqref{bare potential final} can be trusted only when the field dependent mass in the loop integral is sufficiently small: $c_i\varphi^2\ll\Lambda^2$. Nonetheless, let us venture to assume that the expression~\eqref{bare_potential} or \eqref{bare potential final} is still valid even for the field values much larger than the cutoff $\Lambda$.\footnote{
As is suggested in Ref.~\cite{Atick:1988si}, the number of the effective degrees of freedom may be greatly reduced above the string scale. If this is the case, the naive cutoff removing the modes $\ab{p}>\Lambda$ might be a good illustration of the correct picture.
}
When we can take $c_j\varphi^2\gg\Lambda^2$ for some species $j$, say $j=W,Z,t$, and $c_i\varphi^2\ll\Lambda^2$ for others $i$, then
\al{
{\df\V_\text{eff}\over\df\varphi^2}
	&\to
		{1\over2}\sqbr{
			m_B^2
			+\lambda_B^2\varphi^2
			+{\Lambda^4\over16\pi^2\varphi^2}\sum_\text{$j$: $c_j\varphi^2\gg\Lambda^2$}{N_j\over2}
			}.
			\label{high energy limit}
}
We note that the bare mass re-appears in this limit.
As one can see from Fig.~\ref{SMcouplings}, both the bare coupling $\lambda_B$, approximated by the \MSbar one $\lambda(\Lambda)$, and the bare mass~$m_B^2$ are very close to zero for $\Lambda\gtrsim 10^{17}\GeV$. If the UV theory somehow chooses the bare mass to be zero, as is proposed by Veltman, and also $\lambda_B=0$, then the effective potential becomes
\al{
\V_\text{eff}
	&\to	\V_0+{\Lambda^4\over16\pi^2}\ln{\varphi\over\Lambda}\,\sum_\text{$j$: $c_j\varphi^2\gg\Lambda^2$}{N_j\over2}
		\label{high energy limit}
}
at $c_j\varphi^2\gg\Lambda^2$, where $\V_0$ is an integration constant. We see that the potential at very high scales takes the form of Eq.~\eqref{log potential}.

We can read off the coefficient $\V_1$ in Eq.~\eqref{log potential} from Eq.~\eqref{high energy limit}:
\al{
\V_1
	&	
	=	-{3\over32\pi^2}\Lambda^4
}
when we put $j=W,Z,t$. We see that we need to add extra scalar fields coupling to the Higgs to make $\V_1$ positive at high scales, as in the Higgs portal dark matter scenario. As an illustration, we assume hereafter that the Higgs potential is not modified up to the cutoff $\Lambda$ and is connected to Eq.~\eqref{log potential} directly at $\Lambda$ with arbitrary constants $\V_0$ and $\V_1$, though generally the RGE itself can be changed by the inclusion of the extra scalar fields.

\begin{table}[tn]
\begin{center}
$\begin{array}{c|ccccc}
i		&h&\chi&W&Z&t\\
\hline
N_i	&
	1&3&6&3&-4N_c\\
c_i	&3\lambda_B&\lambda_B&{g_{2B}^2/4}&{(g_{2B}^2+g_{YB}^2)/4}&{y_{tB}^2/2}
\end{array}$
\caption{Constants given in Eq.~\protect\eqref{bare_potential} for species $i$ providing large $c_i$.}
\label{number_table}
\end{center}
\end{table}

\section{Limiting behavior}
We show the limiting behavior of the high energy potential Eq.~\eqref{log potential}.
\begin{itemize}
\item In the limit $\V_1\ll\wt\V_0$, we get $\wt\V_0\to \V_0$,
\al{
\epsilon_V
	&\to	{1\over2}\paren{M_P\over\varphi}^2\paren{\V_1\over\V_0}^2,	&
\eta_V
	&\to	-\paren{M_P\over\varphi}^2{\V_1\over\V_0},
		\nn
\xi_V^2
	&\to	2\paren{M_P\over\varphi}^4\paren{\V_1\over\V_0}^2,	&
\varpi_V^3
	&\to	-6\paren{M_P\over\varphi}^6\paren{\V_1\over\V_0}^3,
}
and hence
\al{
{\varphi_\text{end}\over M_P}
	&\to	\sqrt{\V_1\over\V_0}\ll1,	&
N_*	&\to
		{\V_0\over2\V_1}\paren{\varphi_*\over M_P}^2,	&
n_s	&\to
		1-{1\over N_*},	&
r	&\to
		{4\V_1\over N_*\V_0}.
}
For an observed value of $A_s$, we get
\al{
{\V_0\over M_P^4}
	&=	\sqrt{6\pi^2A_s\V_1\over N_*M_P^4}.
}
Or if we remove $\V_1=\V_0\varphi_*^2/2N_*M_P^2$,
\al{
{\V_0\over M_P^4}
	&=	{3\pi^2A_s\over N_*^2}{\varphi_*^2\over M_P^2}.
}
Or else, we can rewrite $\V_1=\paren{\varphi_\text{end}\over M_P}^2\V_0$ to yield
\al{
\V_0
	&=	{6\pi^2A_s\over N_*}\varphi_\text{end}^2M_P^2,	&
\V_1
	&=	{6\pi^2A_s\over N_*}\varphi_\text{end}^4,	&
\varphi_*
	&=	\sqrt{2N_*}\,\varphi_\text{end}.
}
\item In the opposite limit $\wt\V_0\ll\V_1$, we get
\al{
\epsilon_V
	&\to
		{1\over2}\paren{M_P\over\varphi}^2\paren{1\over \ln{\varphi\over M_P}}^2,	&
\eta_V
	&=	-\paren{M_P\over\varphi}^2{1\over \ln{\varphi\over M_P}},
		\nn
\xi_V^2
	&=	2\paren{M_P\over\varphi}^4\paren{1\over \ln{\varphi\over M_P}}^2,	&
\varpi_V^3
	&=	-6\paren{M_P\over\varphi}^6\paren{1\over \ln{\varphi\over M_P}}^3.
}
We define the end point of the inflation by the condition: $\max\br{\epsilon_V,\ab{\eta_V}}=\epsilon_V=1$, to get:
\al{
\varphi_\text{end}
	&=	e^{W(1/\sqrt{2})}M_P
	=	1.57M_P.
}
Then the $e$-folding number becomes
\al{
N_*
	&\to
		-{\varphi_*^2-\varphi_\text{end}^2\over M_P^2}
		+{\varphi_*^2\over 2M_P^2}\ln{\varphi_*\over M_P}
		-{\varphi_\text{end}^2\over 2M_P^2}\ln{\varphi_\text{end}\over M_P},
}
which gives $\varphi_*=8.54M_P$ ($7.96M_P$) for $N_*=60$ (50), and hence
\al{
n_s	&\to	0.994\quad(0.993),	&
r	&\to	6.2\times10^{-3}\quad(7.2\times10^{-3}).
}
This limit $\wt\V_0\ll\V_1$ gives the log-only potential.
Note that
\al{
\V_\text{end}
	&=	W(1/\sqrt{2})\,\V_1
	=	0.45\,\V_1.
}
We also get
\al{
\V_*
	=	\V_1\ln{\varphi_*\over M_P}
	&=	24\pi^2A_s\epsilon_VM_P^4,
}
which gives $\V_1=1.23\times10^{-9}M_P^4$ ($1.46\times10^{-9}M_P^4$) and $\V_\text{end}=3.23\times10^{-9}M_P^4$ ($3.75\times10^{-9}M_P^4$) for $N_*=60$ (50).
\end{itemize}

\bibliographystyle{TitleAndArxiv}
\bibliography{HKO}

\end{document}